\documentclass[aps,pra,twocolumn,superscriptaddress,notitlepage]{revtex4-1}
\usepackage[english]{babel}
\usepackage{graphicx}
\usepackage{upgreek}
\usepackage[linktoc=page,colorlinks,urlcolor=blue,citecolor=blue,linkcolor=blue]{hyperref}
\usepackage{amssymb,amsmath}
\usepackage{graphicx}
\usepackage{natbib}
\usepackage{mathrsfs}
\usepackage{overpic}
\usepackage{wrapfig}
\usepackage{xcolor}
\usepackage{textcomp}
\usepackage{comment}
\usepackage{multirow}
\usepackage{gensymb}
\usepackage[normalem]{ulem}
\usepackage{color}
\usepackage{tabularx}
\usepackage[letterpaper,textwidth=7in,top=.75in,bottom=.75in]{geometry}
\newcommand{\huPhys}{Department of Physics, Harvard University, Cambridge, MA 02138, USA}
\newcommand{\esix}{Element Six Global Innovation Centre, Fermi Avenue, Harwell Oxford, Didcot, Oxfordshire OX11 0QR, United Kingdom.}
\newcommand{\LM}{Lockheed Martin, 199 Borton Landing Road, 101-202 Moorestown, NJ 08057-0927, USA}
\newcommand{\CfA}{Harvard-Smithsonian Center for Astrophysics, Cambridge, MA 02138, USA}
\newcommand{\cbs}{Center for Brain Science, Harvard University, Cambridge, MA 02138, USA}
\newcommand{\MaryPhys} {Department of Physics, University of Maryland, College Park, MD 20740, USA}
\newcommand{\MaryECE} {Department of Computer and Electrical Engineering, University of Maryland, College Park, MD 20740, USA}
\newcommand{\MaryQTC} {Quantum Technology Center, University of Maryland, College Park, MD 20740, USA}
\newcommand{\EPS}{Department of Earth and Planetary Sciences, Harvard University, Cambridge, MA 02138, USA}

\begin{document}

\title{Generation of nitrogen-vacancy ensembles in diamond for quantum sensors: Optimization and scalability of CVD processes}
\date{\today}

\author{Andrew M. Edmonds}\email{andrew.edmonds@e6.com}\affiliation{\esix}
\author{Connor A. Hart} \affiliation{\huPhys}
\author{Matthew J. Turner} \affiliation{\huPhys} \affiliation{\cbs}
\author{Pierre-Olivier Colard} \affiliation{\esix}
\author{Jennifer M. Schloss} \affiliation{\huPhys} \affiliation{\cbs}
\author{Kevin Olsson} \affiliation{\MaryECE}
\author{Raisa Trubko} \affiliation{\huPhys} \affiliation{\EPS}
\author{Matthew L. Markham} \affiliation{\esix}
\author{Adam Rathmill} \affiliation{\esix}
\author{Ben Horne-Smith} \affiliation{\esix}
\author{Wilbur Lew} \affiliation{\LM}
\author{Arul Manickam} \affiliation{\LM}
\author{Scott Bruce} \affiliation{\LM}
\author{Peter G. Kaup} \affiliation{\LM}
\author{Jon C. Russo} \affiliation{\LM}
\author{Michael J. DiMario} \affiliation{\LM}
\author{Joseph T. South} \affiliation{\LM}
\author{Jay T. Hansen} \affiliation{\LM}
\author{Daniel J. Twitchen} \affiliation{\esix}
\author{Ronald L. Walsworth}  \affiliation{\huPhys} \affiliation{\cbs} \affiliation{\MaryECE} \affiliation{\CfA} \affiliation{\MaryPhys} \affiliation{\MaryQTC}

\begin{abstract}
Ensembles of nitrogen-vacancy (NV) centers in diamond are a leading platform for practical quantum sensors. Reproducible and scalable fabrication of NV-ensembles with desired properties is crucial. This work addresses these challenges by developing a chemical vapor deposition (CVD) synthesis process to produce diamond material at scale with improved NV-ensemble properties for a target NV density. The material reported in this work enables immediate sensitivity improvements for current devices. In addition, techniques established in this work for material and sensor characterization at different stages of the CVD synthesis process provide metrics for future efforts targeting other NV densities or sample geometries.
\end{abstract}

\maketitle

\section{Introduction} \label{introduction}

The nitrogen-vacancy (NV) center in diamond is a defect of trigonal ($C_{3v}$) symmetry that has been widely studied over the last decade.  This is by virtue of the fact that the negative-charge state (NV$^\text{-}$) has a spin $S\!=\!1$ ground state that may be initialized and read-out optically~\cite{doherty_nitrogen-vacancy_2013} and coherently controlled through the application of microwaves, with the spin-state having long coherence times even at room-temperature~\cite{balasubramanian_ultralong_2009,stanwix_coherence_2010}.

\begin{figure}[h!]
\begin{center}
\begin{overpic}[width=0.95\columnwidth]{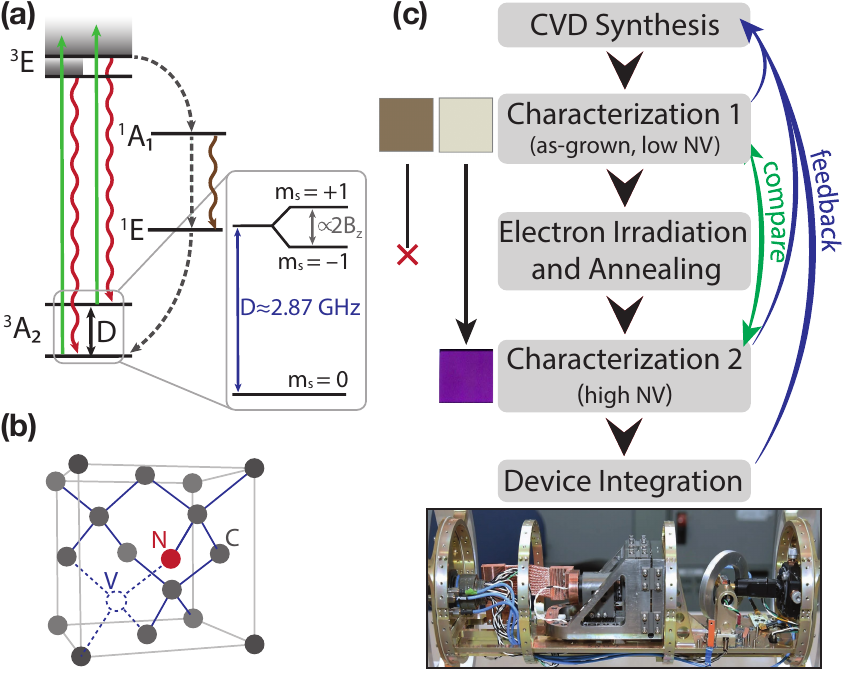}  
\end{overpic}
\end{center}
\caption{NV$^\text{-}$ energy level and structure diagrams and material development procedure. (a) Energy level diagram for the NV$^\text{-}$ center in diamond with zero-field-splitting between ground-state electronic spin levels $m_s\!=\!0$ and $m_s\!=\!\pm1$. Expanded zoom depicts the Zeeman splitting of the $m_s\!=\!\pm1$ energy levels due to an applied magnetic field $B_z$ along the NV symmetry axis. (b) Structural diagram of the nitrogen-vacancy center in diamond. (c) Schematic summarizing the development of a scalable process for producing diamond material optimized for NV-ensemble magnetometry applications. Material is characterized before and after irradiation and annealing; and feedback at each stage informs optimization of the CVD synthesis parameters. Evolution of sample color from a dull brown or yellow to an intense, uniform purple color after irradiation and annealing is a result of high [NV$^\text{-}$]  with minimal unwanted other defects. A device designed for mobile magnetic navigation applications is depicted as an example application.}
\label{fig:diagram}
\end{figure}

The ability to detect and control single NV$^\text{-}$ centers was largely responsible for the initial interest in this color center~\cite{jelezko_single_nodate} and it was established that, in addition to its potential use as a qubit or source of single-photons, diamond containing NV$^\text{-}$ is a useful platform for the detection of electric fields, magnetic fields, temperature, and forces~\cite{taylor_high-sensitivity_2008,degen_scanning_2008,acosta_temperature_2010}. For example, magnetic-field (B) measurements may be made through probing the NV spin levels, which are split by the electronic Zeeman interaction (Fig.~\ref{fig:diagram}(a)), utilizing either DC or AC-detection schemes~\cite{rondin_magnetometry_2014}. For these single NV demonstrations, the availability of suitable high-purity material grown by the Chemical Vapor Deposition (CVD) method was crucial~\cite{Isberg2002,Gibney2014QuantumPF}.

More recently, ensembles of NV$^\text{-}$ centers have been demonstrated to provide routes to high-sensitivity and low-drift broadband B-field sensing, reaching picotesla (or lower) sensitivities under ambient (room temperature) conditions~\cite{wolf_subpicotesla_2015}. This is whilst providing intrinsic vector-field~\cite{maertz_vector_2010,pham_magnetic_2011} measurement capabilities through detection of all four of the NV orientations permitted by the defect’s symmetry in diamond (refer to Fig.~\ref{fig:diagram}(b)). Wide-field B-field imaging using NV-ensembles~\cite{le_sage_optical_2013,glenn_single-cell_2015,shao_wide-field_2016,tetienne_quantum_2017,glenn_micrometer-scale_2017} has enabled diverse applications, including in biology~\cite{le_sage_optical_2013,glenn_single-cell_2015}, geophysics~\cite{glenn_micrometer-scale_2017}, materials science~\cite{tetienne_quantum_2017,simpson_magneto-optical_2016,ku_imaging_2019}, and probing electronic circuits~\cite{jfRochICs, ICpaper}. NV-ensemble magnetometry also has potential applications in RF-sensing~\cite{chipaux_wide_2015}, magnetic navigation~\cite{canciani_absolute_2016}, magnetic-anomaly detection~\cite{sheinker_magnetic_2009}, and geo-surveying. Consequently there is now industrial interest in utilizing NV-ensemble sensors, with prototype devices constructed and being used outside the lab~\cite{sturner_compact_2019,webb_nanotesla_2019}. An example is the mobile magnetic navigation application depicted in Fig.~\ref{fig:diagram}(c). Crucial to the success of these efforts is the wide availability of diamond samples offering ensembles of NV centers with reproducible properties, at a range of well-controlled concentrations appropriately chosen for each application and its practical constraints (size, weight, power, sensitivity, etc.).

\subsection{Material considerations for optimal NV magnetic field sensitivity} \label{materialconsiderations}
For NV-ensembles, the optical shot noise limited DC magnetic sensitivity ($\eta$) is given by~\cite{budker_optical_2007,acosta_diamonds_2009}:

\begin{equation}
\eta \sim \frac{1}{g_e \mu_B } \frac{1}{C\sqrt{\beta}}  \frac{1}{\sqrt{N_\text{NV} T_2^*}}
\label{sensitivity}
\end{equation}

where $N_\text{NV}$ is the number of NV$^\text{-}$ centers that are utilized in the sensor (given by the product of the concentration of NV$^\text{-}$ centers, [NV$^\text{-}$], and the interrogated volume of the diamond), $T_2^*$ is the ensemble spin dephasing time, $\beta$ is the optical detection efficiency, and $C$ is the measurement contrast. The physical constants $g_e$ and $\mu_B$ are the Landé factor and Bohr magneton, respectively. Consequently, the sensitivity is both a function of the diamond material and the overall sensor design. The material-related factors $N_\text{NV}$, $C$, and $T_2^*$ have recently been the topic of an in-depth review on routes to optimize sensitivity~\cite{barry_sensitivity_2019}, which further motivates the work presented in this paper. 

A typical approach to create NV centers in diamond is to start with a sample produced by high-pressure high-temperature (HPHT) or CVD synthesis containing substitutional nitrogen (N$_\text{S}$); to electron-irradiate to create vacancies (V); and then to anneal at temperatures $>\!600\,^\circ$C, where the V are mobile (see Fig.~\ref{fig:diagram}(c) for example images of material at different stages)~\cite{davies_vacancy-related_1992}. It should be noted that NV centers are typically found in as-grown CVD diamond~\cite{kennedy_long_nodate} unless considerable efforts are made to exclude nitrogen from the chamber, but are only present as a small fraction of the overall N-content~\cite{edmonds_production_2012}. The negative-charge state NV$^\text{-}$, which has the physical properties utilized in sensing, arises from the donation of an electron (typically from N$_\text{S}^0$) according to NV$^0 + \text{N}_\text{S}^0 \rightarrow \text{NV}^\text{-} + \text{N}_\text{S}^+$. The neutral charge state NV$^0$ exhibits an optical luminescence spectrum that overlaps with that of NV$^\text{-}$; thus NV$^0$ will contribute to the background luminescence in a typical device, degrading the contrast $C$. As a result, it is important to consider the concentration of both NV$^0$ and NV$^\text{-}$ in a sample from the perspective of increasing contrast ($C$). Therefore the fraction of NV$^\text{-}$ becomes an important figure of merit:

\begin{equation}
\psi= \frac{[\text{NV}^\text{-}]}{[\text{NV}^\text{-}]+[\text{NV}^0]}	
\label{eq:chargefraction}
\end{equation}

Factors influencing $\psi$ include the starting level of [N$_\text{S}$] in the diamond material, which acts as an upper limit of the possible level of [NV]; the irradiation dose (i.e., [V]); and the annealing recipe used to convert N$_\text{S}$ and V present post-irradiation into NV. Other defects, X, present in the diamond, either post-growth (CVD-specific examples are discussed in Sec.~\ref{CVDdiamondapps}) or post-irradiation~\cite{lawson_existence_nodate}, may additionally act as donors/acceptors and influence $\psi$.

[NV$^\text{-}$], [NV$^\text{0}$] and [N$_\text{S}^0$] also influence the resulting ensemble NV dephasing time $T_2^*$, as they contribute to the electronic spin-bath. [X] is also a potential factor in determining $T_2^*$, if the defects are paramagnetic. $^{13}$C has a nuclear spin of $I\!=\!\frac{1}{2}$ and therefore adds to the nuclear spin-bath. It is thus typical to produce diamond samples with depleted levels of [$^{13}$C] in order to maximize $T_2^*$~\cite{balasubramanian_ultralong_2009}. The final source of ensemble dephasing intrinsic to the diamond material is non-uniform strain across the area of the diamond sample being utilized~\cite{dolde_electric-field_2011,kehayias_imaging_2019-1}. Considering these contributions, the material-related $T_2^*$ can be approximated by the following expression~\cite{bauch_ultralong_2018,barry_sensitivity_2019}:

\begin{equation}
\begin{split}
\frac{1}{T_2^*(\text{material})} \approx \frac{1}{T_2^*(\text{N}_\text{S}^0)}+\frac{1}{T_2^*(\text{NV}^\text{-})}\\
+\frac{1}{T_2^*(\text{NV}^0)}+\frac{1}{T_2^*(\text{X})}+\frac{1}{T_2^*(^{13}\text{C})}+\frac{1}{T_2^*(\text{strain})}
\end{split}
\label{eq:t2starmaterial}
\end{equation}
Since the $N_\text{NV}$ term in Eq.~\ref{sensitivity} is given by the product of [NV$^\text{-}$] and the interrogated volume of the diamond, these two variables are important characteristics to examine for material and sensor design. In particular, increasing [NV$^\text{-}$] can lead to reduced $T_2^*$, e.g., through increased [N$_\text{S}$] (Eq.~\ref{eq:t2starmaterial}). Thus, a critical figure of merit is the product of [NV$^\text{-}$] and $T_2^*$. However, since [N$_\text{S}^0$] limits the level of [NV$^\text{-}$] that can be produced by irradiation and annealing, a key material-related decision is the starting [N$_\text{S}$]. The interplay between the optimal [NV$^\text{-}$] and $T_2^*$ at different [$^{13}$C], and resulting consequences for magnetic sensitivity, can therefore be informed by considering the effect of [N$_\text{S}^0$] and [$^{13}$C] on $T_2^*$.

\begin{figure}[htbp]
\begin{center}
\begin{overpic}[width=0.95\columnwidth]{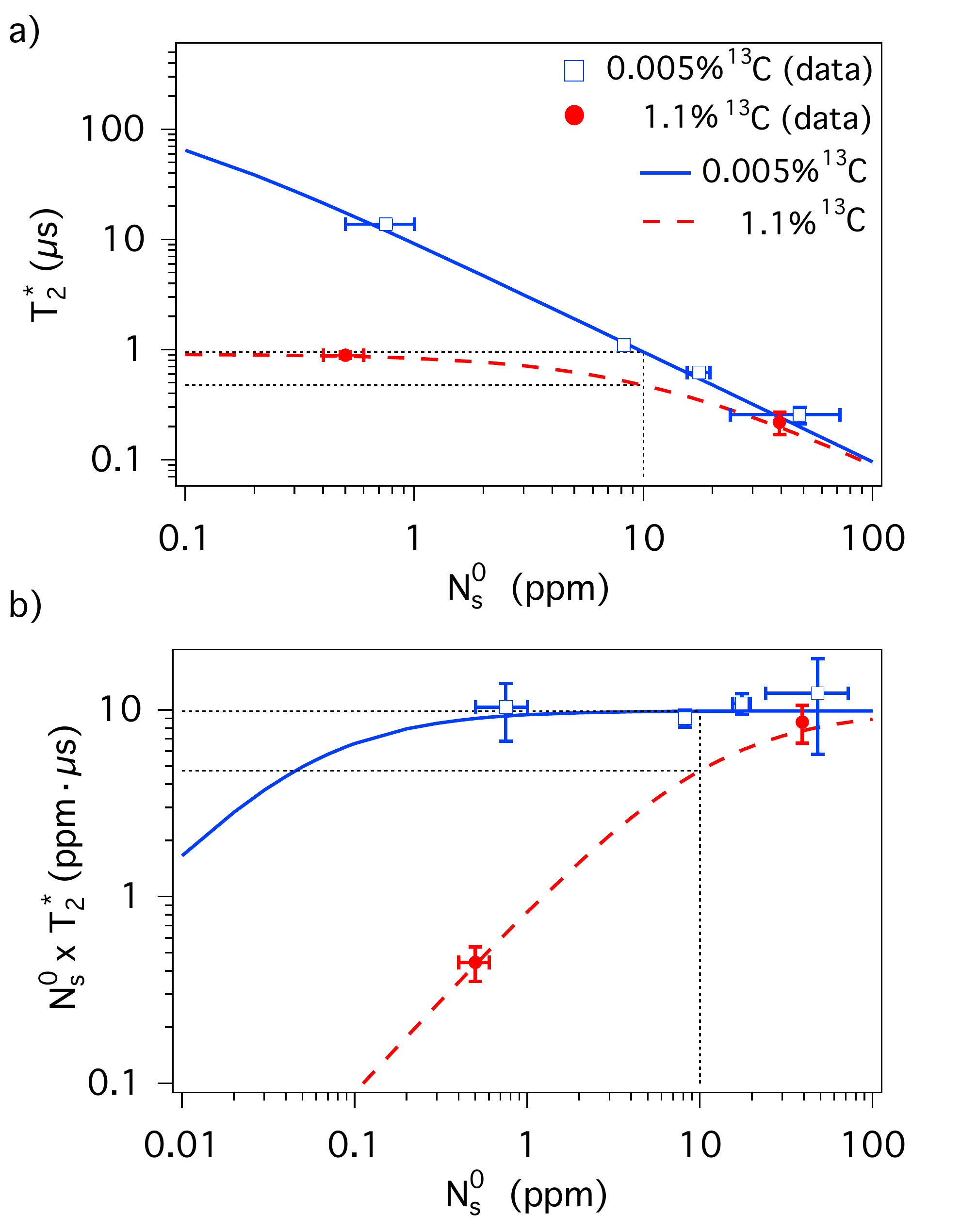}
\end{overpic}
\end{center}
\caption{\label{fig:FOM}
(a) Expected dependence of ensemble NV$^\text{-}$ $T_2^*$ with varying [N$_\text{S}^0$] and [$^{13}$C], according to Eq.~\ref{eq:t2s13CNs0}. (b) Product of [N$_\text{S}^0$] and $T_2^*$ as a function of [N$_\text{S}^0$]. The [N$_\text{S}^0$]$\,\sim\,$10-15$\,$ppm  regime, the focus of the present paper, is indicated by the dashed, black lines. Comparisons to representative measurements from our collaboration are shown~\protect \cite{bauch_ultralong_2018,barry_sensitivity_2019}.}
\end{figure}

The expected $T_2^*$, assuming that [N$_\text{S}^0$] and [$^{13}$C] are the dominant contributors to the dephasing time, can be estimated using the expression:

\begin{equation}
\frac{1}{T_2^*(^{13}\text{C},\text{N}_\text{S}^0)}\approx{A_{^{13}\text{C}}\times[^{13}\text{C}]}+{A_{\text{N}_\text{S}^0}\times[\text{N}_\text{S}^0]}
\label{eq:t2s13CNs0}
\end{equation}
where, from previous measurements, $A_{^{13}\text{C}}\!\approx\!0.100\,$ms$^{-1}$ppm$^{-1}$ and $A_{\text{N}_\text{S}^0}\!\approx\!101\,$ms$^{-1}$ppm$^{-1}$~\cite{barry_sensitivity_2019,bauch_ultralong_2018}. The contribution from NV$^\text{-}$ centers after irradiation and annealing is expected to be proportional to [N$_\text{S}^0$] and is thus not explicitly included in the subsequent analysis. The dependence of $T_2^*$ on [N$_\text{S}^0$] for natural abundance $^{13}$C (1.1$\%$, 11000$\,$ppm) and depleted $^{13}$C (0.005$\%$, 50$\,$ppm) is illustrated in Fig.~\ref{fig:FOM}(a), including example measurements performed by our collaboration and reported in past work~\cite{bauch_ultralong_2018,barry_sensitivity_2019}.

Fig.~\ref{fig:FOM}(b) depicts the product of [N$_\text{S}^0$] and $T_2^*$ for both natural abundance $^{13}$C (1.1$\%$, 11000$\,$ppm) and isotopically depleted $^{13}$C (0.005$\%$, 50$\,$ppm). Across the range of [N$_\text{S}^0$] depicted, the concentration of $^{13}$C ([$^{13}$C]) has a critical role in determining both the achievable magnetic sensitivity and optimal [N$_\text{S}^0$]. 

From Fig.~\ref{fig:FOM}(a,b) it is apparent that for [N$_\text{S}^0$] below $\sim\,$100$\,$ppm, $^{13}$C-isotopic depletion is advantageous, extending $T_2^*$ and increasing the figure of merit [N$_\text{S}^0$]$\,\times\,T_2^*$. Furthermore, as discussed in Barry et al., if nitrogen-related dephasing is a small contribution to $T_2^*$, then the nitrogen concentration should be increased until similar to the dominant dephasing source~\cite{barry_sensitivity_2019}. This is illustrated by the plateau in the product of [N$_\text{S}^0$] and $T_2^*$ for increasing [N$_\text{S}^0$] in Fig.~\ref{fig:FOM}(b). While in natural abundance material [N$_\text{S}^0$]$\,\times\,T_2^*$ plateaus at [N$_\text{S}^0$]$\,\sim\!100\,$ppm, $^{13}$C-depletion reduces the optimal [N$_\text{S}^0$] to approximately 1-20$\,$ppm. 

Since the figure of merit for $^{13}$C-depleted material is largely constant in the range 1-20$\,$ppm, additional factors should be considered when choosing a target [N$_\text{S}^0$] in this regime. For pulsed magnetometry protocols such as Ramsey, lower [N$_\text{S}^0$] and longer $T_2^*$ may provide advantages such as improved measurement duty cycle. However, achieving longer $T_2^*$ in practice requires better control of other dephasing sources such as magnetic bias field gradients and strain inhomogeneity across an interrogated NV$^\text{-}$ ensemble. Consequently, [N$_\text{S}^0$] of order 10-15$\,$ppm is attractive because it relaxes these material and sensor design requirements without degrading the figure of merit [N$_\text{S}^0$]$\,\times\, T_2^*$. These considerations are especially critical when increasing the sensing volume for bulk magnetometry and wide-field magnetic field imaging applications using NV-ensembles. 

With these considerations, the present work focuses on material with [N$_\text{S}^0$]$\,\sim\,$10-15$\,$ppm. As outlined in the following section, this nitrogen concentration has traditionally proven a challenging regime in which to produce, by CVD growth, favorable diamond material for NV magnetic-field-sensing applications.  It is also generally recognized that the fabrication of high-[NV] samples by CVD with reasonable coherence properties is non-trivial and such material is suggested as a near-term target in a recent review~\cite{achard_2020} by Achard et al.

In this work, samples with both natural abundance and isotopically depleted $^{13}$C concentrations are compared to aid development of an efficient framework for process optimization of future material with other target [N$_\text{S}^0$] using less expensive natural abundance $^{13}$C methane.

\subsection{CVD diamond optimization for ensemble NV magnetic-field sensors} \label{CVDdiamondapps}

The present work focuses on CVD diamond, which is applicable to the widest range of NV sensing modalities and applications. Due to the morphology that evolves during growth, HPHT single-crystal diamond has different sectors present (e.g., $\{100\}$ and $\{111\}$) and the incorporation of N differs significantly between these regions~\cite{Burns1990}. Hence HPHT material must be processed into plates consisting of a single-sector, which is challenging and can limit the volume of any given sensor element that can be produced. Also, it remains unclear whether controllable and reproducible levels of [N$_\text{S}$] and hence [NV$^\text{-}$] can be obtained in HPHT diamond. Additionally, it is not practical with the HPHT method to create diamonds with NV-ensemble surface layers for wide-field magnetic-imaging  applications.  

Here, we address two key challenges for the optimization of CVD methodologies that are essential for the production of diamond for NV-ensemble magnetometry applications. Firstly, a limitation of CVD synthesis of N-containing diamond is the incorporation of additional, undesired defects. In particular, diamond grown by the CVD method may exhibit a brown coloration, with strong correlations observed between the N concentration in the process gases during CVD growth (necessary to produce N$_\text{S}$ in the material) and the level of  broadband-absorption features that give rise to this brown color~\cite{martineau_identification_2004,tallaire_highly_2017}. Such features are thought to arise from vacancy chains and clusters~\cite{khan_color_2010, khan_charge_2009, hounsome_origin_2006, fujita_large_2009} that are incorporated during synthesis. Such defects act as traps for electrons~\cite{khan_charge_2009, campbell_lattice_2002} and hence can reduce the NV charge fraction $\psi$. Associated with this effect, N-doped CVD diamond can contain significant [N$_\text{S}^+$]~\cite{edmonds_production_2012,khan_colour-causing_2013}. These defects, as well as other commonly-observed H-related defects in CVD diamond (e.g., the nitrogen-vacancy-hydrogen defect, NVH~\cite{glover_hydrogen_nodate}), exhibit charge states that are paramagnetic~\cite{glover_hydrogen_nodate, glover_hydrogen_nodate-1} and thus act as a source of dephasing, contributing to the $1/(T_2^* (X) )$ term in Eq.~\ref{eq:t2starmaterial}. Minimizing such parasitic defects and understanding the links between material characteristics and charge fraction is thus a key challenge. Secondly, increasing [N] to the tens of ppm level in the CVD process gases has been observed to promote the formation of extended or non-epitaxial defects during growth~\cite{willems_optical_2014,tallaire_growth_2013}, thereby making it challenging to realize high-[N] material with homogeneous strain~\cite{gaukroger_x-ray_2008,friel_control_2009} and thus spatially-uniform and long $T_2^*$.

This paper reports a study of N-doped CVD processes across a range of synthesis conditions (Sec.~\ref{samplesynthesis}), examining the resulting broadband-optical-absorption characteristics and level of charge acceptors in as-grown CVD material (Sec.~\ref{asgrown}). The goal is to optimize production of samples containing $\sim\,$10-15$\,$ppm [N$_\text{S}^0$], while inhibiting the level of parasitic defects and strain heterogeneity and increasing the NV$^\text{-}$ charge-state fraction $\psi$ and spin-state readout contrast. Irradiated and annealed samples are then studied to assess the influences of strain and parasitic defects on key metrics relevant for ensemble NV sensors: [NV$^\text{-}$], $\psi$, and $T_2^*$ (Sec.~\ref{postirradanneal}). Based on these studies, prospects for reproducibly producing samples with controlled levels of strain, [NV], and $T_2^*$ are then discussed (Sec.~\ref{strainbatch}). Finally, in Sec.~\ref{parametersimpact} the interplay between CVD processes and NV$^\text{-}$ charge-state and spin-state readout contrast are explored.

\section{Sample synthesis, treatment and characterization methods} \label{samplesynthesis}

The samples examined in this paper were produced by CVD in a microwave-plasma-assisted reactor. $\{100\}$-oriented single-crystal CVD diamond plates containing [N$_\text{S}^0$]$\,\sim\,$0.1$\,$ppm acted as substrates during each deposition run. A range of synthesis conditions were utilized, in order to produce batches of samples with varying levels of [N$_\text{S}$] and optical absorption characteristics. This process consisted of a wide range of synthesis conditions, covering variations in substrate temperature, T$_\text{sub}\!\approx\,$800‒1100$\,^\circ$C, N concentration in the gas phase (N$_\text{gas}\!\approx\,$10‒150$\,$ppm), and methane fraction, CH$_4$/H$_2\!\approx\,$1-5$\,\%$. CH$_4$ sources were used that either had natural abundances of C-isotopes, or were enriched to 99.995$\,\%$ $^{12}$C. Synthesis was stopped once the diamond layer thickness reached $\sim\,$500$\,$-$\,$1000$\,$\micro m in each run, in order to permit the use of multiple characterization techniques to examine the [N$_\text{S}$], optical absorption, and strain of the grown material.

The resulting samples were irradiated using an electron beam energy of 4.5\,MeV whilst placed on a water-cooled metal bench. At this beam energy, the electron dose would be expected to be homogeneous through the thicknesses of samples grown for this paper (few hundred microns)~\cite{campbell_lattice_2002}. Samples were irradiated for varying durations with the electron dose then estimated from the geometry of the system and the known current of the e- source. 

Subsequent annealing of the samples to create NV centers took place in a tube furnace with samples placed in an alumina boat. After loading, the tube was evacuated to a pressure of $\sim\,$1$\times$10$^{\text{-}6}$ mbar in order to minimize graphitization. Annealing was undertaken with the following thermal-ramp profile: 400$\,^\circ$C for 2 hours, 800$\,^\circ$C for 16 hours, 1000$\,^\circ$C for 2 hours and 1200$\,^\circ$C for 2 hours (3$\,^\circ$C/min ramp rate), similar to the methodology employed by Chu et al~\cite{chu_coherent_2014}.

Room-temperature optical absorption measurements to probe the absorption characteristics of samples in the range 240-800$\,$nm (UV-Vis) were performed using an Analytik Jena Specord 50 Plus spectrometer. This permitted measurement of [N$_\text{S}^0$] and estimates of the strength of absorption band features at 360 and 520\,nm through spectral deconvolution and fitting of the samples post-synthesis as described by Khan et al.~\cite{khan_colour-causing_2013}. Fourier Transform Infrared spectroscopy (FTIR) spectroscopy was also used to estimate [N$_\text{S}^0$] as well as [N$_\text{S}^+$] in the as-grown samples, through measurement and fitting of the absorption peaks at 1130$\,$cm$^{\text{-}1}$ and 1344$\,$cm$^{\text{-}1}$ for N$_\text{S}^0$ and 1332$\,$cm$^{\text{-}1}$ for N$_\text{S}^+$ [28]; see Liggins for further details~\cite{liggins_identication_2010}. These techniques employed an aperture of 1.5$\,$mm.

Electron Paramagnetic Resonance (EPR) at X-band frequencies ($\sim\,$9.7 GHz) was used in order to quantify the level of paramagnetic defects NVH$^\text{-}$, N$_\text{S}^0$ and NV$^\text{-}$ in samples prior to irradiation and annealing.  A sample of known [N$_\text{S}^0$] was used as a reference and the spectral fitting and deconvolution method is described elsewhere~\cite{edmonds_magnetic_2008,tallaire_characterisation_nodate}.

Irradiated and annealed samples were examined by low-temperature (77\,K) UV-Vis absorption measurements, with samples held within an Oxford Instruments Optistat DN2 cryostat and cooled to 77\,K using liquid N$_2$. The integrated intensities under the zero-phonon-lines (ZPLs) at 575$\,$nm and 637$\,$nm were then used to quantify the levels of [NV$^0$] and [NV$^\text{-}$] respectively, using the revised calibration constants of Dale~\cite{dale_colour_2015} (updated from those of Davies~\cite{davies_current_1999}). Prior to quantification of defect concentrations by the methods described, samples were exposed to UV for 2 minutes, using the Xe arc lamp excitation source of the DiamondView photoluminescence imaging instrument~\cite{martineau_identification_2004}.

NV-based characterization of diamond material produced in this work employed two setups. The first setup was designed for wide-field continuous wave optically detected magnetic resonance (CW-ODMR) imaging of mm-scale diamond samples as previously described in Ref.~\cite{kehayias_imaging_2019-1}. From the measured CW-ODMR spectra in each pixel, both magnetic and strain-induced shifts in the NV$^\text{-}$ spin resonances were determined by fitting to the NV$^\text{-}$ Hamiltonian as described in Ref.~\cite{glenn_micrometer-scale_2017,kehayias_imaging_2019-1}.

The second photodiode-based setup utilized pulsed microwave control to measure the NV$^\text{-}$ ensemble $T_2^*$ by extracting the Ramsey free induction decay constant. Using an epi-illumination microscope configuration, 5$\,$-$\,$1000$\,$mW of 532$\,$nm laser light were focused through the sample with a beam-waist of 20$\,$\micro m. A 2$\,$mT applied bias magnetic field aligned with NV$^\text{-}$ centers oriented along a single crystallographic axis induced a Zeeman splitting such that the $m_s\!=\!0$ to $m_s\!=\!\pm1$ transitions between the NV$^\text{-}$ ground state sublevels were individually addressable with resonant MW pulses. The applied bias field homogeneity was previously engineered to ensure negligible contributions to $T_2^*$ for the samples measured in this work~\cite{bauch_ultralong_2018}. 
Ramsey-based measurements enabled determination of $T_2^*$ for both the single and double quantum coherences. For double quantum (DQ) Ramsey measurements (immune to axial strain-induced contributions to $T_2^*$), two-tone MW pulses resonant with the NV$^\text{-}$ ground state spin transitions prepared a superposition of the $m_s\!=\!\pm1$ states during the free precession interval. Single quantum (SQ) Ramsey measurements (sensitive to axial strain-induced contributions to $T_2^*$) employed single-tone MW pulses to create a superposition of the $m_s\!=\!0$ and $m_s\!=\!+1$ or $m_s\!=\!-1$ levels during the free precession interval. See Sec.~\ref{strainmitigation} and Bauch et al.~\cite{bauch_ultralong_2018} for further discussion of single and double quantum coherence measurements. Alternatively a CW-ODMR linewidth ($\gamma$) measurement was used as a proxy for the single quantum $T_2^*=1/(\pi\gamma)$, as it is compatible with batch analysis of samples (see Supplemental Material~\cite{suppl})

Quantitative birefringence microscopy was used to assess the level of strain in samples after laser cutting and polishing of the surfaces.  This was performed using a commercial Metripol system~\cite{glazer_automatic_1996}, with the methodology as discussed by Friel et al.~\cite{friel_control_2009}.  Images were collected through the growth face of the sample, since dislocations that thread in the growth direction are the dominant contribution to strain in CVD diamond~\cite{martineau_identification_2004, gaukroger_x-ray_2008, friel_control_2009}.

\section{Process optimization using as-grown material} \label{asgrown}

This section describes the characterization of samples across the range of growth conditions outlined in Sec.~\ref{samplesynthesis}, including their UV-Vis absorption properties, resulting color, and concentration of [N$_\text{S}^0$] and [N$_\text{S}^+$]. This was undertaken to elucidate growth conditions that are likely to be beneficial to produce irradiated and annealed samples with $\sim\,$ppm levels of [NV] and desirable properties for B-field sensing, since defects responsible for color are likely to impact the spin and electronic properties of CVD diamond. Samples of a preferred synthesis recipe are then further characterized by EPR. These initial studies were conducted with natural-abundance CH$_4$ gas (98.9$\%$ $^{12}$C, 1.1$\%$ $^{13}$C).

\subsection{Characterizing the nitrogen and charge environment with processes P$_1$ and P$_2$} \label{absorptionetc}

\begin{figure}[b]
\begin{center}
\begin{overpic}[width=0.95\columnwidth]{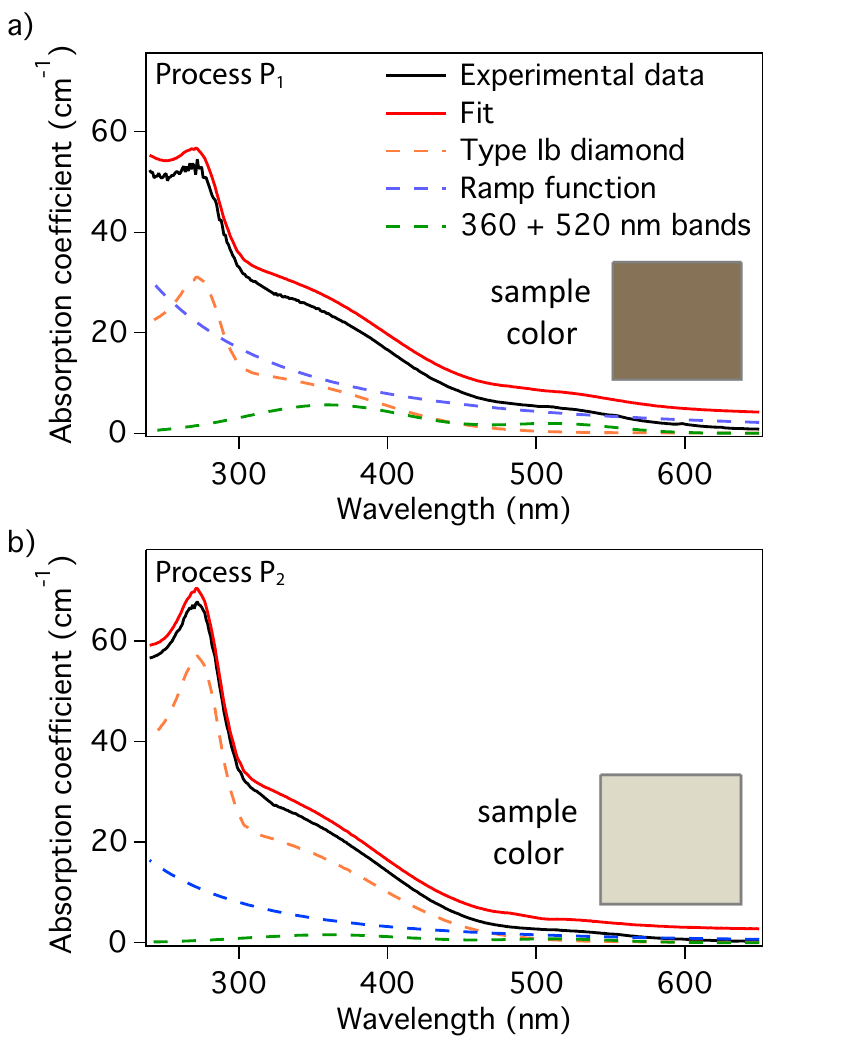}
\end{overpic}
\end{center}
\caption{\label{fig:P1P2UVVis}
Representative UV-Vis absorption spectra of as-grown high-[N] CVD material. (a) Material grown using a CVD process yielding $\approx\,$9.3$\,$ppm [N$_\text{S}^0$] and $\approx\,$13$\,$ppm [N$_\text{S}$]. (b) Material grown with a CVD process yielding $\approx\,$17.0$\,$ppm [N$_\text{S}^0$]. In both plots, the fit has been displaced vertically for clarity and an indication of the sample color is shown.}
\end{figure}

\begin{figure}[h]
\begin{center}
\begin{overpic}[width=0.95\columnwidth]{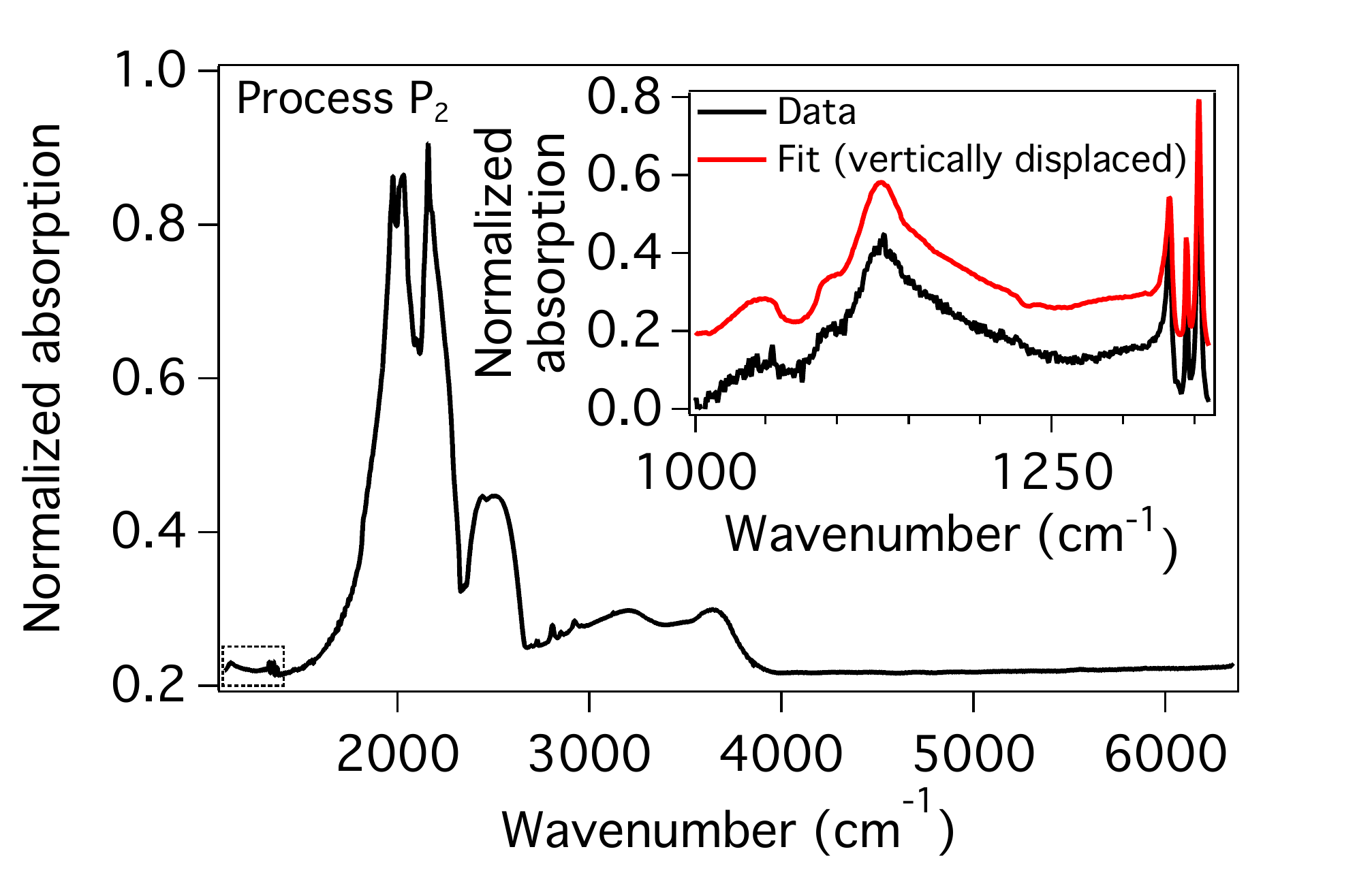}
\end{overpic}
\end{center}
\caption{\label{fig:P2FTIR}
Representative FTIR spectrum of synthesized as-grown high-[N] CVD material for process P$_2$. The inset shows an enlargement of the region of the spectrum showing the 1130$\,$cm$^{-1}$ and 1344$\,$cm$^{-1}$ peaks (associated with N$_\text{S}^0$) as well as the 1332$\,$cm$^{-1}$ peak (associated with N$_\text{S}^+$). The fit has been displaced vertically for clarity.}
\end{figure}

A wide range of levels of [N$_\text{S}$] and absorption characteristics were observed as the synthesis conditions were altered. Example UV-Vis absorption and FTIR spectra, used to quantify [N$_\text{S}^0$] and [N$_\text{S}^+$] are shown in Fig.~\ref{fig:P1P2UVVis} and Fig.~\ref{fig:P2FTIR}, respectively.

\begin{table}[b]
\centering
\caption{Results from high-[N] diamond samples after CVD growth for two different processes (P$_1$ and P$_2$), illustrating the difference in nitrogen concentrations ([N$_\text{S}^0$] and [N$_\text{S}^+$]) and color of the samples as evaluated by lightness (L$^*$). Quoted results are averages across 5 samples in each run. Example UV-Vis absorption and FTIR spectra are shown in Fig.~\ref{fig:P1P2UVVis} and Fig.~\ref{fig:P2FTIR}, respectively.}
\label{tab:p1p2comparison}
\renewcommand{\arraystretch}{1.3}
\renewcommand{\tabcolsep}{8pt}
\begin{tabular}{ccccc}
\hline \hline
\textbf{Process} & \textbf{\begin{tabular}[c]{@{}c@{}}{[}N$_\text{S}^0${]} \\ (ppm)\end{tabular}} & \textbf{\begin{tabular}[c]{@{}c@{}}{[}N$_\text{S}^+${]} \\ (ppm)\end{tabular}} & \textbf{\begin{tabular}[c]{@{}c@{}}{[}N$_\text{S}^0${]}/{[}N$_\text{S}${]} \\ ($\chi$)\end{tabular}} & \textbf{L$^*$} \\ \hline
P$_1$ & 9.3$\,$(9) & 3.5$\,$(7) & 0.73$\,$(3) & 52$\,$(1) \\
P$_2$ & 17$\,$(1) & 3.0$\,$(3) & 0.85$\,$(1) & 74$\,$(2) \\ \hline
\end{tabular}%
\end{table}

In the case of the UV-Vis absorption spectra, a peak at 270$\,$nm was observed, which is attributed to N$_\text{S}^0$~\cite{dyer_optical_1965,chrenko_dispersed_1971} as well as bands at 360$\,$nm and 520$\,$nm, which are thought to originate from clusters of vacancies and NVH$^0$ respectively~\cite{khan_charge_2009}. In addition, a ramp in absorption as wavelength is decreased (of the form $\lambda^{-3}$) was present, as discussed in previous studies~\cite{khan_charge_2009,khan_colour-causing_2013}. These components in the overall absorption spectra are shown in Fig.~\ref{fig:P1P2UVVis}. In the case of the “type Ib” diamond component, an HPHT sample of known concentration was used and scaled appropriately to fit the spectrum and determine [N$_\text{S}^0$]. Prior to characterization, the samples were exposed to UV, as described in Sec.~\ref{samplesynthesis}. This ensures samples are in a consistent state prior to measurement and has the effect of maximizing [N$_\text{S}^0$] (and minimizing [N$_\text{S}^+$]~\cite{khan_colour-causing_2013}).

To illustrate the spread of results, samples produced by two processes (P$_1$ and P$_2$) at the extreme ends of the conditions examined are reviewed in Table~\ref{tab:p1p2comparison}. One process yielded samples with a total [N$_\text{S}$] (given by the sum of [N$_\text{S}^0$] and [N$_\text{S}^+$]) of 13$\,$(1)$\,$ppm, the second produced samples with [N$_\text{S}$] = 20$\,$(1)$\,$ppm, where the uncertainty reflects the variation between the samples in each run.  This demonstrates that within each deposition (set of synthesis conditions) the samples were reasonably consistent, but across these two processes dramatic differences in [N$_\text{S}^0$] and [N$_\text{S}^+$] were observed. 

It was apparent that the color of the two sets of samples differed significantly (refer to Fig.~\ref{fig:P1P2UVVis}). In order to assess this in a quantitative manner, images of the samples were examined in ImageJ~\cite{schneider_nih_2012} after the background was normalized to pure white, given by a lightness (L$^*$) value of 100 (CIELAB color space~\cite{mclaren1976}) where lightness indicates the relative brightness of a color (an L$^*$ value of 0 corresponds to pure black). The color was averaged over a circular area in the center of the samples and the L$^*$ value for each sample was determined; in this sense L$^*$ was used as a proxy for the degree of brown coloration in the sample. The values of L$^*$ determined and shown in Table~\ref{tab:p1p2comparison} illustrate that the samples from process P$_2$ were lighter in color (lower brown). Hence, through careful choice of synthesis conditions, higher levels of nitrogen doping in CVD processes do not necessarily cause a higher degree of brown coloration in as-grown samples.

It was also notable that the fraction of [N$_\text{S}^0$]/[N$_\text{S}$] (defined as $\chi$) significantly differed between the two sets of samples. This suggests that the fraction of defects acting as charge acceptors is different between the two groups. Prior studies of CVD material over a range of N-doping levels have also demonstrated how this charge fraction can vary; in other samples containing [N$_\text{S}$]$>$10$\,$ppm it was observed that [N$_\text{S}^0$]$\,\approx\,$[N$_\text{S}^+$]~\cite{edmonds_production_2012}. The lowest value determined for $\chi$ in this current study was similar, measuring $\sim\,$0.5, for samples containing $\sim\,$17$\,$ppm [N$_\text{S}$].

It has previously been suggested in studies by Khan et al.~that the presence of high levels of [N$_\text{S}^+$] in CVD diamond are indicative of significant brown color~\cite{khan_colour-causing_2013}. Hence this was investigated across the entire range of explored synthesis conditions to elucidate any correlations that may exist between the charge fraction of N$_\text{S}$, the color of the samples and the absorption features present in spectra (such as those presented in Fig.~\ref{fig:P1P2UVVis}).  

\begin{figure}[htbp]
\begin{center}
\begin{overpic}[width=0.95\columnwidth]{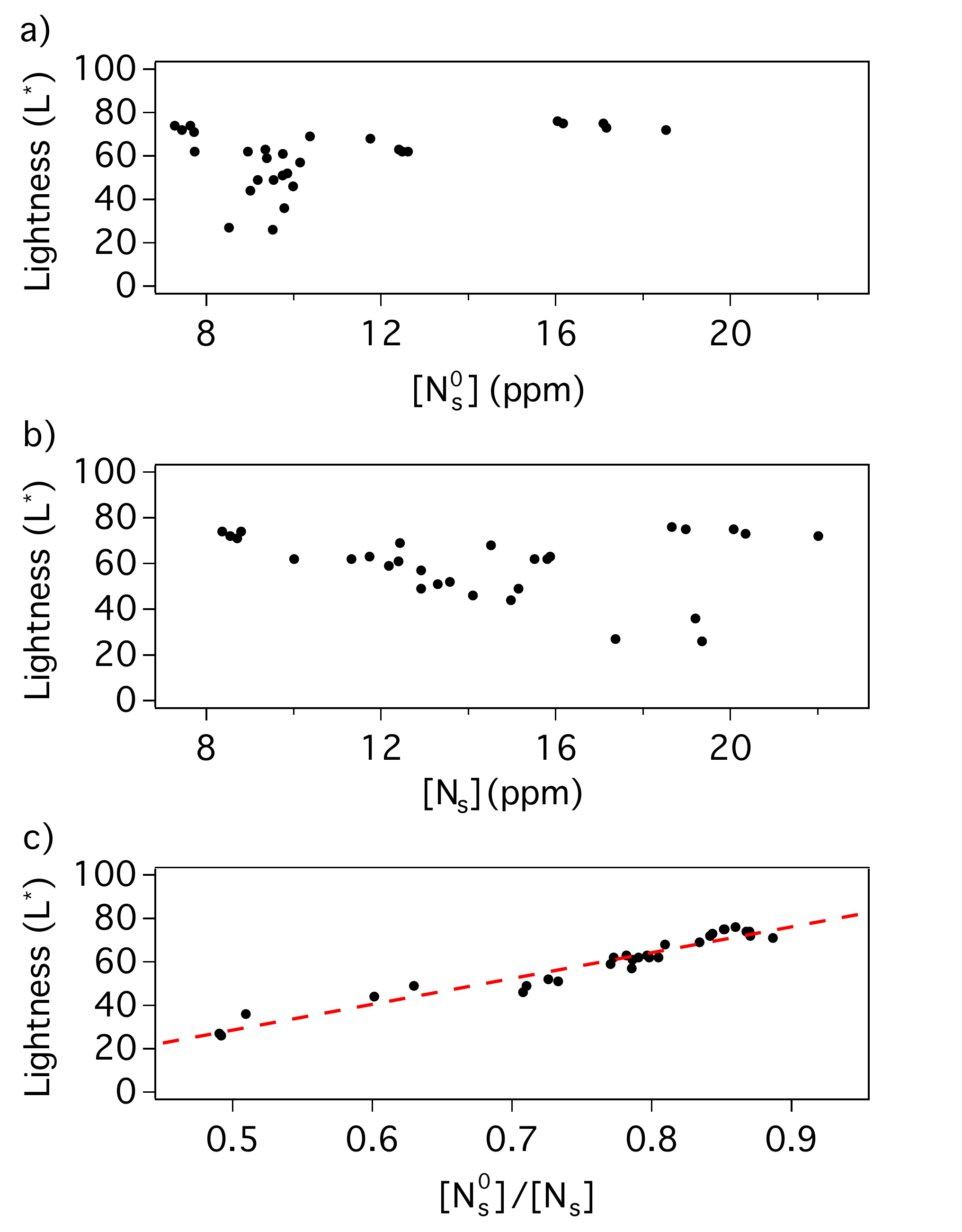}
\end{overpic}
\end{center}
\caption{\label{fig:Lightness}
(a) For all material produced in this study, plot of sample lightness (evaluated as L$^*$) as function of neutral substitutional nitrogen concentration, [N$_\text{S}^0$]. (b) Plot constructed using the same data set as (a), but as a function of total [N$_\text{S}$] (summing [N$_\text{S}^0$] and [N$_\text{S}^+$]) and (c) L$^*$ against the ratio of [N$_\text{S}^0$] to [N$_\text{S}$] (denoted by $\chi$ in the text). The linear fit in (c) is a guide to the eye to illustrate the link between the two parameters.}
\end{figure}

Fig.~\ref{fig:Lightness} illustrates the data from all diamond material produced in this study. The lightness of the as-grown samples was observed to be correlated not with the determined [N$_\text{S}^0$] in the samples, nor with the total [N$_\text{S}$], but with the charge fraction $\chi$, [N$_\text{S}^0$]/[N$_\text{S}$]. This suggests that the level of absorption leading to brown coloration is associated with the degree of acceptor-defects present in the material, consistent with the comments by Khan et al.~\cite{khan_colour-causing_2013}.

\begin{figure}[htbp]
\begin{center}
\begin{overpic}[width=0.95\columnwidth]{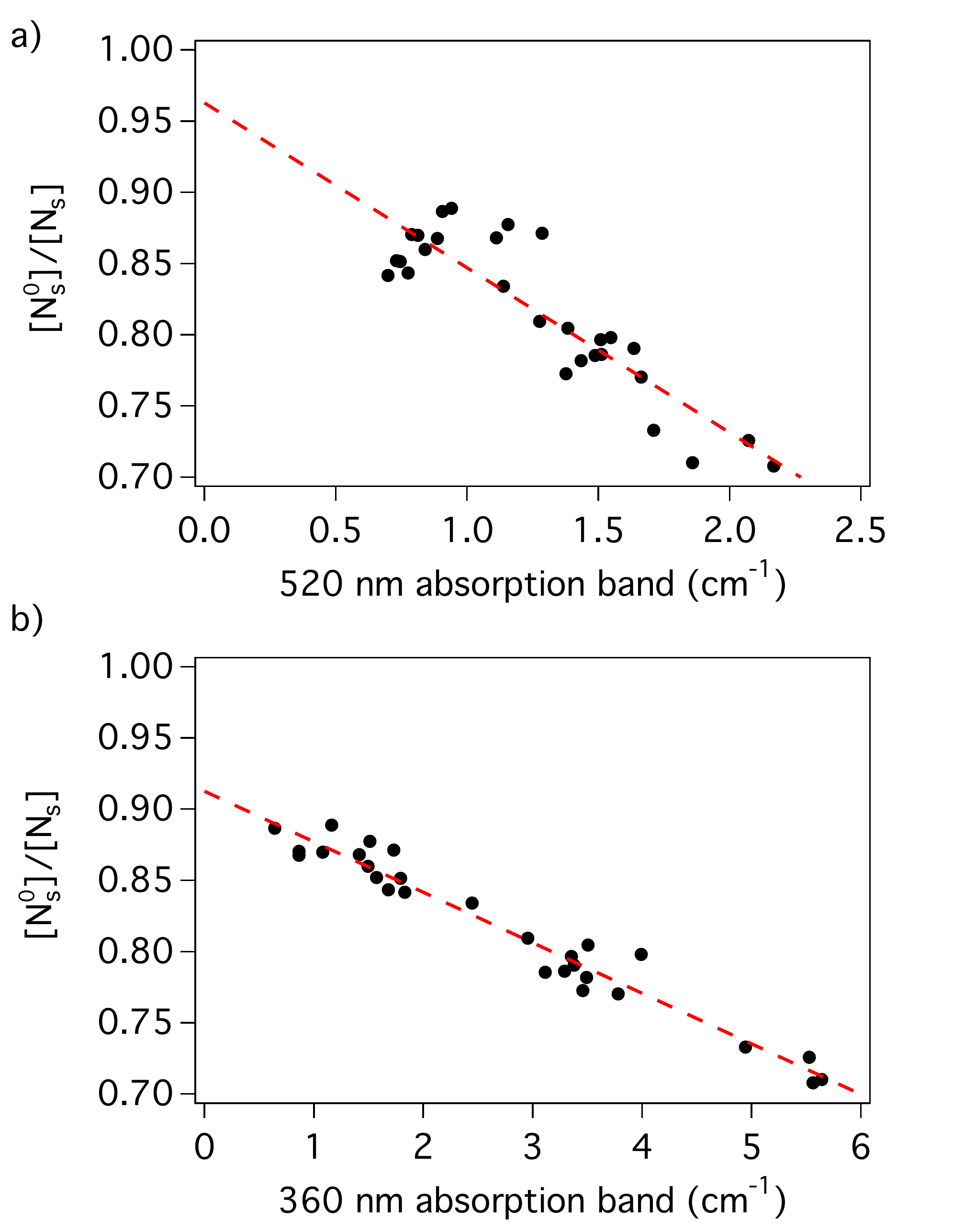}
\end{overpic}
\end{center}
\caption{\label{fig:ChargeNbands}
For all material produced in this study, relationship  between the charge fraction [N$_\text{S}^0$]/[N$_\text{S}$] (denoted by $\chi$ in the text) and the strength of (a) the 520$\,$nm absorption band and (b) the 360$\,$nm absorption band observed in an absorption spectrum measurement. Linear fit is a guide to the eye.}
\end{figure}

Fig.~\ref{fig:ChargeNbands} illustrates the relationship between the strength of the 520$\,$nm and 360$\,$nm bands and the determined N$_\text{S}$ charge-fraction value. In both cases an increase in the absorption feature was attributed to a decrease in $\chi$, establishing that the defects responsible for these bands are acceptors. A higher spread in the results for the 520$\,$nm band is evident (Fig.~\ref{fig:ChargeNbands}(a)), but is likely due to the relative weakness of this feature in the spectra. The examples shown in Fig.~\ref{fig:P1P2UVVis} act as a good demonstration of this behavior, as the 360$\,$nm and 520$\,$nm bands are approximately a factor of three lower in Fig.~\ref{fig:P1P2UVVis}(b) compared to Fig.~\ref{fig:P1P2UVVis}(a), which correlates with the difference in charge-state fraction and color of the samples, as shown in Table~\ref{tab:p1p2comparison}.  

It is also noteworthy that, if a linear relationship is assumed between the N$_\text{S}$ charge fraction and the strength of the absorption feature in both plots in Fig.~\ref{fig:ChargeNbands}, the line of best-fit trends to unity in the absence of the band. Hence, if the defects causing the 360$\,$nm and 520$\,$nm features were absent, negligible N$_\text{S}^+$ would be expected. N$_\text{S}^+$ is generally not observed in untreated HPHT samples and neither are these two bands.

\begin{table*}[htbp]
\caption{Summary of as-grown N-related defect concentrations in a representative sample from the high-[N] CVD process P$_3$ as measured by UV-Vis and FTIR absorption spectroscopy as well as electron paramagnetic resonance (EPR).}
\label{tab:p3characterization}
\renewcommand{\arraystretch}{1.25}
\renewcommand{\tabcolsep}{8pt}
\begin{tabular}{ccccclll}
\hline \hline
\multicolumn{3}{c|}{\textbf{\begin{tabular}[c]{@{}c@{}}{[}N$_\text{S}^0${]}\\ (ppm)\end{tabular}}} &
  \multicolumn{1}{c|}{\textbf{\begin{tabular}[c]{@{}c@{}}{[}N$_\text{S}^+${]}\\ (ppm)\end{tabular}}} &
  \multicolumn{1}{c|}{\textbf{\begin{tabular}[c]{@{}c@{}}{[}N$_\text{S}${]}\\ (ppm)\end{tabular}}} &
  \multicolumn{1}{c|}{\textbf{\begin{tabular}[c]{@{}c@{}}{[}NVH$^-${]}\\ (ppm)\end{tabular}}} &
  \multicolumn{2}{c}{\textbf{\begin{tabular}[c]{@{}c@{}}{[}NV$^-${]}\\ (ppm)\end{tabular}}} \\ \hline
UV-Vis &
  FTIR &
  \multicolumn{1}{c|}{EPR} &
  \multicolumn{1}{c|}{FTIR} &
  \multicolumn{1}{c|}{FTIR} &
  \multicolumn{1}{c|}{EPR} &
  EPR &
  UV-Vis \\
13.9$\,$(7) &
  15$\,$(2) &
  \multicolumn{1}{c|}{16$\,$(2)} &
  \multicolumn{1}{c|}{3.5$\,$(7)} &
  \multicolumn{1}{c|}{19$\,$(1)} &
  \multicolumn{1}{c|}{1.6$\,$(2)} &
  0.08$\,$(1) &
  0.070$\,$(4) \\ \hline
 &
   &
   &
   &
   &
  \multicolumn{1}{l}{} &
  \multicolumn{1}{l}{} &
  \multicolumn{1}{l}{}
\end{tabular}
\end{table*}

The mapping from CVD process values such as the flow of H, CH$_4$ and other gases, dopant level, and substrate temperature T$_\text{sub}$ onto the physical parameters controlling growth, e.g., the density of surface radical sites, the rate of $\text{CH}_x$ addition relative to etching by H atoms, and the near-surface $\text{NH}_x$ (or CN)/$\text{CH}_x$ ratio, depends on the particular reactor design via intermediate variables such as the gas and electron temperatures and the position of the plasma relative to the deposition area. The process conditions needed to produce given material characteristics therefore differ considerably between different reactor designs. Nevertheless, in this study it was established that, at a given doping level, careful simultaneous control of the CH$_4$ fraction (relative to total gas flow) and the substrate temperature  was crucial to reducing $\chi$ whilst also maintaining a growth surface free of etch pits~\cite{achard_control_2005} or $\{100\}$ surface twins~\cite{wild_oriented_1994}.

Following the findings outlined in this section, a simple metric for a desirable CVD recipe is high [N$_\text{S}^0$] and a low degree of brown-coloration. These two properties were readily measurable in as-grown samples with no material processing required, as the effect of a substrate with $\sim\,$0.1$\,$ppm [N$_\text{S}^0$] is negligible in terms of color and measured [N$_\text{S}^0$]. Hence, traversing  a range of growth conditions and rapidly characterizing the material was possible.

\subsection{Development and characterization of process P$_3$} \label{hightanneal}

Following arguments presented in Sec.~\ref{materialconsiderations} a process targeting [N$_\text{S}^0$] $\sim15$\,ppm whilst maintaining a high fraction of [N$_\text{S}^0$]/[N$_\text{S}$] was desired.  Based on the findings from the samples characterized in Sec.~\ref{absorptionetc}, a process denoted P$_3$ was developed. A small initial batch of 5 diamond samples resulted in  [N$_\text{S}^0$]$\,\approx\!14\,(1)\,$ppm and [N$_\text{S}^0$]/[N$_\text{S}$]$\,\approx0.81\,(2)$, and were utilized for further characterization and processing. 

EPR measurements were conducted on a sample from this first batch in order to investigate additional point defects present in this material. This approach allowed [NVH$^\text{-}$] and [NV$^\text{-}$] to be quantified in samples grown using process P$_3$ prior to treatment. Table~\ref{tab:p3characterization} shows these results and summarizes the quantification of [N$_\text{S}^0$] by three different techniques; UV-Vis and FTIR absorption measurements, as well as EPR, in order to confirm general agreement between these methods. 

The concentrations [NV$^\text{-}$] and [NVH$^\text{-}$] can be compared to [N$_\text{S}$] in order to assess the ratios of N-related defects in this material. [N$_\text{S}$]:[NVH$^\text{-}$]:[NV$^\text{-}$] in the examined sample was $\sim\,$230:20:1, close to the previously observed values in studies of CVD material (300:30:1~\cite{edmonds_production_2012} and 52:7:1~\cite{hartland_study_2014}). Hence, despite having high-[N] and a low fraction of acceptors (high $\chi$), NVH remains a considerable fraction of the measurable N-related defects in the studied material ($>$10$\,\%$, given only the negative charge-state can be quantified). This likely reflects the hydrogen-rich environment that exists during the CVD growth process. 

Motivated by previous reports~\cite{Orwa2011,tetienne_spin_2018,osterkamp_engineering_2019} concerning the annealing of NV$^\text{-}$ containing material at high temperatures, similar experiments were conducted on the P$_3$ samples. Annealing took place in vacuum at 1500$^\circ$C for 16 hrs to maximize any possible effects of treatments at this temperature. As shown in Table~\ref{tab:1500deganneal}, [NV$^\text{-}$] increased to $>\,$0.2$\,$ppm after annealing, suggesting some residual vacancy clusters were broken up in this treatment. [NV$^0$] was below detection limits both before and after annealing ($<$0.01$\,$ppm). A straightforward N$\rightarrow$NV conversion is likely, echoing recent findings in treatment of layers grown on $\{111\}$-oriented substrates~\cite{osterkamp_engineering_2019}. In the results shown in Table~\ref{tab:1500deganneal}, it is also notable that the 360$\,$nm absorption band decreased dramatically in strength (by $\sim\,$90$\,\%$), lending support to previous assignments of V-related defects/clusters to this feature. The 520$\,$nm feature remained unchanged, within the likely uncertainties of the measurements. Further investigation is needed to map out the extent to which vacancy related defects (especially those associated with the 360 nm line) impact NV creation and contribute to spin bath dephasing of NV$^\text{-}$ ensembles~\cite{bauch_ultralong_2018}.

\begin{table}[h]
\caption{Concentrations of N$_\text{S}^0$ and NV$^\text{-}$ as measured by UV-Vis and UV-Vis absorption spectra coefficients at 360$\,$nm and 520$\,$nm before and after sample annealing at 1500$\,^\circ$C for a representative sample from process P$_3$.}
\label{tab:1500deganneal}
\renewcommand{\arraystretch}{1.2}
\renewcommand{\tabcolsep}{6pt}
\begin{tabular}{ccccc}
\hline \hline
 &
  \textbf{\begin{tabular}[c]{@{}c@{}}{[}N$_\text{S}^0${]}\\ (ppm)\end{tabular}} &
  \textbf{\begin{tabular}[c]{@{}c@{}}{[}NV$^\text{-}${]}\\ (ppm)\end{tabular}} &
  \textbf{\begin{tabular}[c]{@{}c@{}}360 nm\\ (cm$^{-1}$)\end{tabular}} &
  \textbf{\begin{tabular}[c]{@{}c@{}}520 nm\\ (cm$^{-1}$)\end{tabular}} \\ \hline
As-grown             & 13.9$\,$(7)           & 0.070$\,$(4)       & 3.0$\,$(1)            & 1.5$\,$(1)            \\
Post-anneal    & 13.7$\,$(7)          & 0.20$\,$(1)       & 0.3$\,$(1)            & 1.7$\,$(1)          \\ \hline
\multicolumn{1}{l}{} & \multicolumn{1}{l}{} & \multicolumn{1}{l}{} & \multicolumn{1}{l}{} & \multicolumn{1}{l}{}
\end{tabular}
\end{table}

\section{Characterization of material post irradiation and annealing and batch analysis} \label{postirradanneal}

This section discusses characterization of diamond samples synthesized using process P$_3$ after being electron irradiated and annealed to create $\sim\,$ppm levels of NV centers. Measurements of [NV], [NV$^\text{-}$], and [NV$^\text{0}$] as a function of electron irradiation dose up to 6$\times$10$^{18}\,$cm$^{-2}$ are presented in Sec.~\ref{NVirraddose}. Utilizing a selected dose, batches of samples are characterized to evaluate the production of material with reproducible [NV$^\text{-}$] at scale in Sec.~\ref{strainbatch}. For these synthesis runs, the success of additional steps before and during growth to mitigate lattice strain inhomogeneity are assessed using birefringence imaging.

\subsection{Nitrogen-vacancy concentration as function of irradiation dose} \label{NVirraddose}

\begin{figure}[htbp]
\begin{center}
\begin{overpic}[width=0.9\columnwidth]{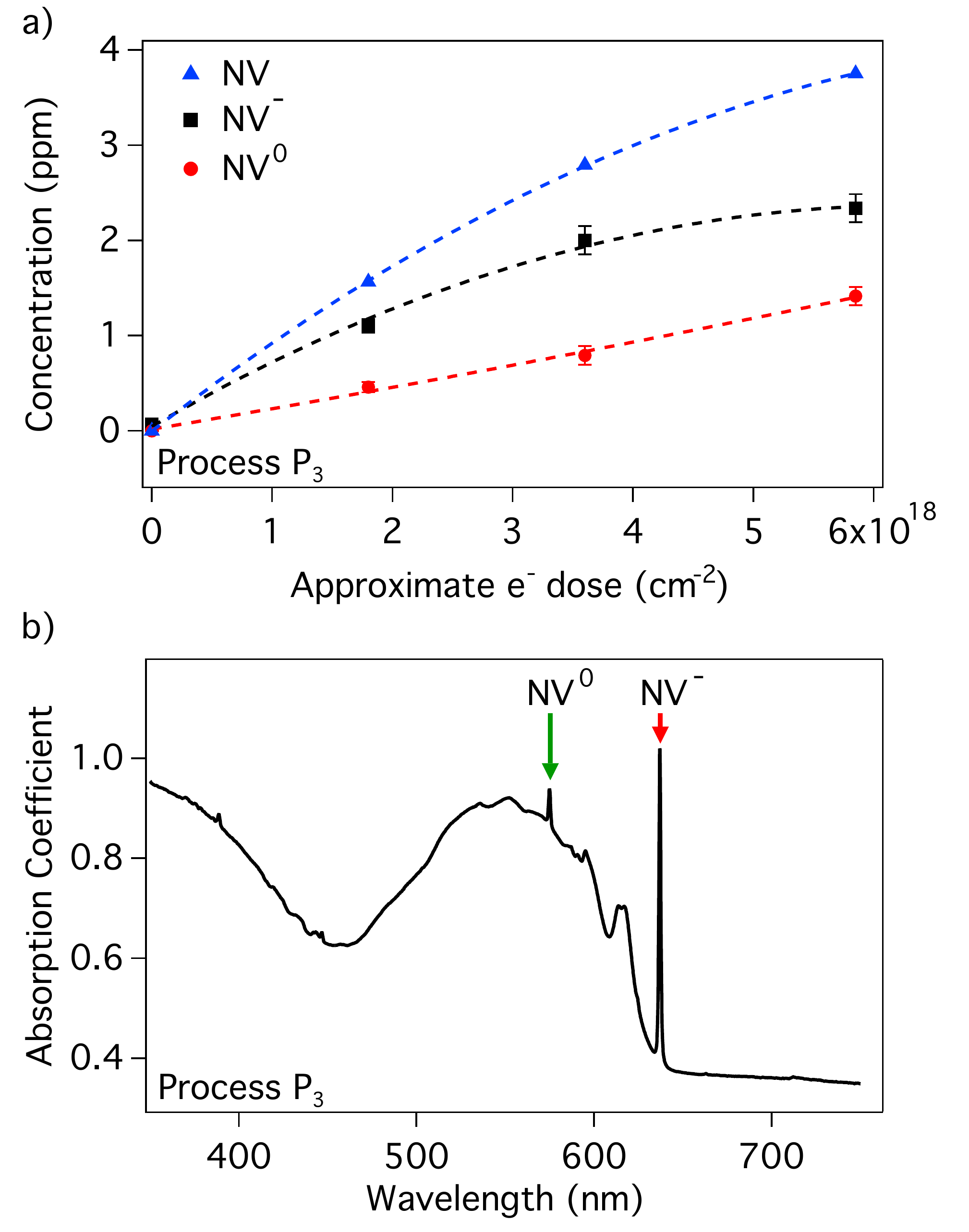}
\end{overpic}
\end{center}
\caption{\label{fig:NVirrad}
(a) Average concentrations of [NV$^\text{-}$], [NV$^0$] and [NV] ([NV$^\text{-}$]+[NV$^0$]) of e$^\text{-}$ irradiated (4.5\,MeV) and annealed process P$_3$ samples containing $\approx$14$\,$ppm [N$_\text{S}^0$] as-grown, as measured by UV-Vis absorption. (b) Example 77 K UV-Vis measurement of process P$_3$ material after irradiation to a e$^\text{-}$ dose of $\sim$6$\times$10$^{18}\,$cm$^{\text{-}2}$ and annealing up to 1200$\,^\circ$C. Measurements were made after UV exposure.}
\end{figure}

To optimize the fraction of [NV$^\text{-}$] ($\psi$) in the material, it is crucial to choose the irradiation dose appropriately. If the irradiation dose and hence number of vacancies introduced is too low, then the generation of NV centers will be limited. Conversely, if the material is over-irradiated, [NV] will be saturated, but at the expense of generating a large number of NV$^0$ centers, detrimentally affecting the value of $\psi$~\cite{mita_change_nodate, waldermann_creating_2007}.

For the process P$_3$ material described in the previous section with $\approx\,$14$\,$ppm [N$_\text{S}^0$] it was therefore desirable to investigate the generation of NV$^\text{-}$ and NV$^0$ centers as a function of irradiation dose. This was conducted up to a dose of $\sim$6$\times$10$^{18}\,$cm$^{-2}$ and the results obtained after the samples were annealed are shown in Fig.~\ref{fig:NVirrad}(a). It should be noted that the samples were annealed using a ramped-temperature annealing recipe which has a final 2 hour step at 1200$\,^\circ$C. Although temperatures above 1000$\,^\circ$C do not increase [NV]~\cite{acosta_diamonds_2009}, higher temperatures have previously been shown to assist in annealing out multi-vacancy defects~\cite{naydenov_increasing_2010,yamamoto_extending_2013}. Measurements of [NV$^\text{-}$] and [NV$^0$] were conducted by UV-Vis after the samples were exposed to UV.

Over this range of irradiation doses [NV$^0$] was observed to increase linearly, whereas [NV$^\text{-}$] began to saturate at the highest dose. The dose was therefore not increased further and was chosen as the level of irradiation to utilize for the remainder of this study. At this chosen level of irradiation, samples were found to contain 3.7$\,$(2)$\,$ppm [NV], comprising 2.3$\,$(1)$\,$ppm of [NV$^\text{-}$] and 1.4$\,$(1)$\,$ppm of [NV$^0$] after exposure to UV. An example 77 K UV-Vis absorption spectrum of this material is shown in Fig.~\ref{fig:NVirrad}(b).

\subsection{Batch analysis of process P$_3$}\label{strainbatch}
 
\begin{figure*}[htbp]
\begin{center}
\begin{overpic}[width=0.95\textwidth]{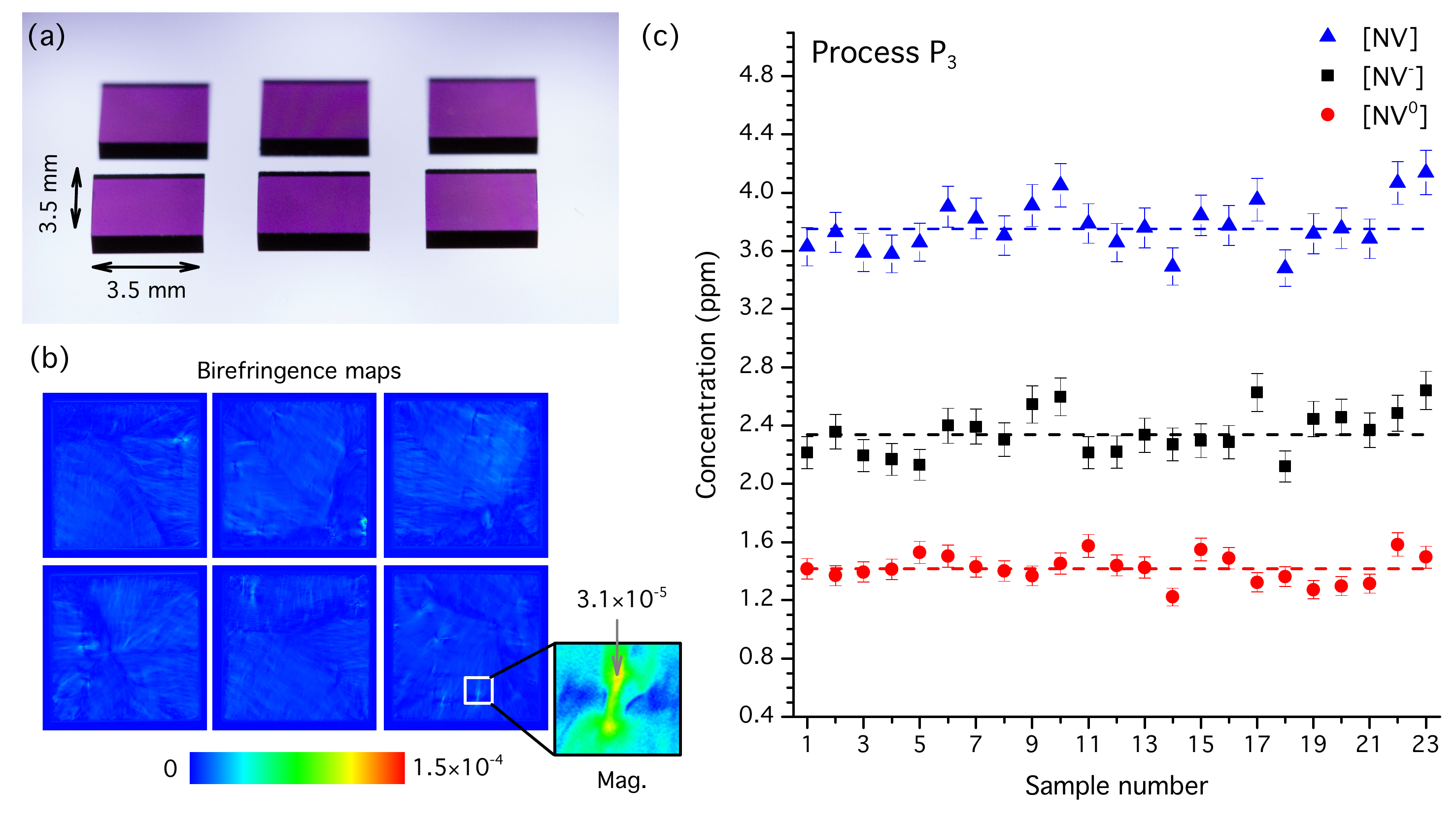}
\end{overpic}
\end{center}
\caption{\label{fig:P3batch}
(a) Photograph of 6 process P$_3$ samples (0.005$\%$ $^{13}$C) with approximately 900 \micro m thickness, after irradiation and annealing to create $\approx\,$3.8$\,$ppm [NV]. (b) Birefringence images of the plates in (a) as measured on a Metripol microscope. (c) Concentration results, determined by UV-Vis absorption, of [NV$^0$], [NV$^\text{-}$] and total [NV] (sum of [NV$^0$] and [NV$^\text{-}$]) across a batch of 23 process P$_3$ samples.}
\end{figure*}

In this section, batch characterization of samples is reported to demonstrate the reproducible production of material with well-controlled [NV$^\text{-}$] and strain inhomogeneity. Samples were synthesized using process P$_3$ (0.005\% $^{13}$C) in combination with additional strain-mitigation techniques: CVD substrates and pre-synthesis etches were carefully controlled to minimize the density of dislocations present in the high-[N] material grown, according to the methods discussed in Friel et al.~\cite{friel_control_2009}. Deposition conditions were controlled for the duration of the run to avoid the formation of non-epitaxial crystallites. 

As-grown samples were irradiated and annealed using the selected dose of 6$\times$10$^{18}\,$cm$^{-2}$ identified previously in Sec.~\ref{NVirraddose}. Examples of these samples, post irradiation and annealing, are shown in Fig.~\ref{fig:P3batch}(a). The intense purple color is a result of the high [NV$^\text{-}$] achieved in this material. Across such a batch of 23 samples, the average [N$_\text{S}^0$] was approximately $13\,$ppm with a standard deviation of 1$\,$ppm, which demonstrates the ability to achieve the same level of [N$_\text{S}^0$] in a larger run, as well as repeatability between separate synthesis runs (refer to Table~\ref{tab:p3characterization} and Table~\ref{tab:1500deganneal}). As shown in Fig.~\ref{fig:P3batch}(c), the measured [NV]$\,=\,$3.8$\,(2)\,$ppm ([NV$^\text{-}$]$\,=\,$2.3$\,$(2)$\,$ppm) was similarly consistent across the batch and with previous samples grown using process P$_3$ (Sec.~\ref{NVirraddose}). The measured [NV$^\text{-}$] and [NV$^\text{0}$] yield a favorable average charge fraction of $\psi=0.62\,(5)$ (uncertainty indicates one standard deviation). 

Reducing the strain inhomogeneity in samples is critical to avoid degrading $T_2^*$ and limiting the magnetic sensitivity of an NV-ensemble device. The strain environment of each sample in the batch was characterized using Metripol birefringence imaging and representative images are shown in Fig~\ref{fig:P3batch}(b). In these samples, an average birefringence $\Delta$n $\approx$ 7$\,$(1)$\times 10^{-6}$ was determined with peak values of $\Delta$n $\sim\,$3$\times10^{-5}$ in isolated petal features (see inset of Fig.~\ref{fig:P3batch}(b) for an example). A vast majority ($>$99$\%$) of the pixel values within the birefringence image, Fig.~\ref{fig:P3batch}(b), satisfy $\Delta$n $\lesssim\,10^{-5}$, the standard for ultra-low birefringence established by Friel et al.~\cite{friel_control_2009}. In the following section, NV-based measurements are reported, demonstrating that the achieved level of strain control is sufficient to avoid limiting NV-ensemble magnetic sensitivity.

\section{Impact on NV Sensing Parameters} \label{parametersimpact}

A representative sample from the previous section was selected for further NV-based characterization. Here,  the properties of this sample relevant to magnetic sensitivity are reported, including $T_2^*$ and ODMR contrast. Correlations between the final material properties and as-grown material properties (N$_\text{S}$ charge fraction, $\chi$) are established by comparing process P$_3$ to a fourth process (P$_4$) with similar [N$_\text{S}$] but dramatically lower charge fraction due to an increased concentration of parasitic defects.

\subsection{Strain mitigation and $T_2^*$ measurements}\label{strainmitigation}
 
 \begin{figure}[htbp]
\begin{center}
\begin{overpic}[width=1.0\columnwidth]{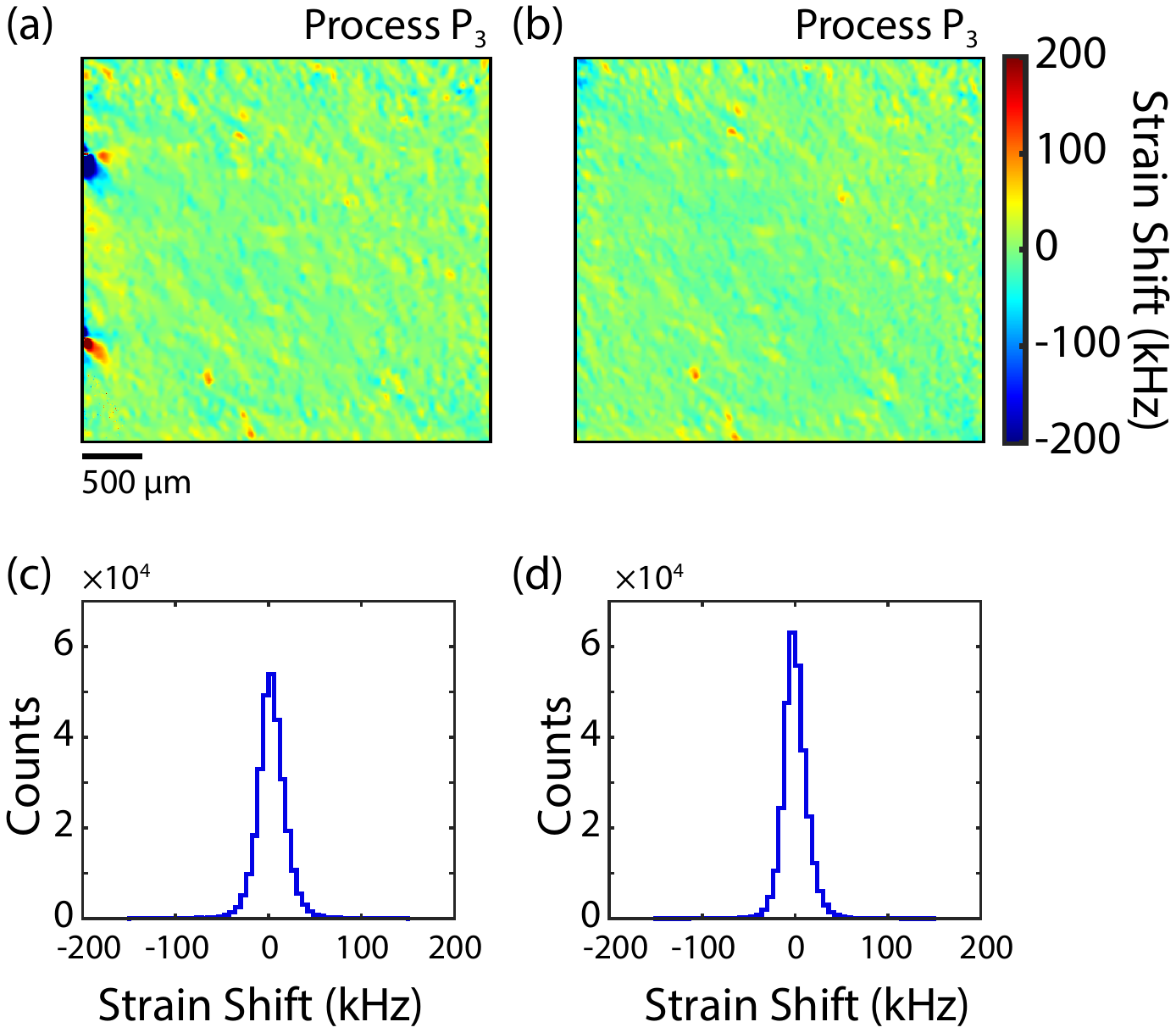}
\end{overpic}
\end{center}
\caption{\label{fig:MzMap}
(a) Map of extracted strain-induced NV resonance shifts for a (3.6$\times$3.6$\times$0.1)$\,$mm$^3$ freestanding plate produced from a thicker original process P$_3$ sample. (b) Map of extracted strain-induced NV resonance shifts for a second freestanding plate produced from a different portion of the sample used to produce the plate shown in (a). (c,d) Histograms of the strain shift values shown in (a) and (b), respectively.}
\end{figure}

Strain in the host diamond lattice induces shifts in the NV$^\text{-}$ spin resonances. When averaging over an ensemble of NV$^\text{-}$ centers, strain inhomogeneity can degrade the ensemble-NV dephasing time, $T_2^*$, and ODMR contrast~\cite{kehayias_imaging_2019-1} (see Supplemental Material~\cite{suppl}). To further characterize the strain mitigation strategies introduced in Sec.~\ref{strainbatch}, the NV$^\text{-}$ spin properties were studied. These measurements employed two 100$\,$\micro m thick freestanding plates produced from one of the ($3.6\times3.6\times0.9)\,$mm$^3$ process P$_3$ samples used in the batch analysis. This thickness was chosen to improve the planar spatial resolution of CW-OMDR-based imaging and reduce inhomogeneities in the applied magnetic, optical, and MW control fields.

Strain-induced resonance shifts were extracted by fitting the measured CW-ODMR spectra pixel-by-pixel to the full NV Hamiltonian as described in Ref.~\cite{glenn_micrometer-scale_2017, kehayias_imaging_2019-1}. Due to the thickness of the diamond substrate limiting spatial resolution,~\cite{glenn_micrometer-scale_2017, kehayias_imaging_2019-1} NV strain shift measurements are only advantageous for probing the strain environment for length scales larger than the thickness of the diamond, 100$\,$\micro m. Maps of these shifts for the two samples are shown in Fig.~\ref{fig:MzMap}(a,b) and histograms of the measured shifts are shown in Fig.~\ref{fig:MzMap}(c,d), respectively. Both samples exhibit minimal strain inhomogeneity with a distribution in strain-induced shifts of approximately 25$\,$kHz full-width half-maximum (FWHM). These measurements demonstrate dramatically improved strain control in the CVD process relative to previous samples in the literature (see Ref.~\cite{kehayias_imaging_2019-1} for typical examples of N-doped CVD diamond with strain-induced shifts on the order of hundreds of kHz to MHz). The shared spatial variations in the strain inhomogeneity between Fig.~\ref{fig:MzMap}(a) and~\ref{fig:MzMap}(b) are a consequence of the two samples being cut from the same source material (particularly visible along the bottom edges of the diamond plates).

For the sample shown in Fig.~\ref{fig:MzMap}(a), additional photodiode-based Ramsey measurements of ensemble-NV $T_2^*$ were conducted on the setup described in Sec.~\ref{samplesynthesis} and previously established to have negligible contribution from $B_0$ gradients and temporal variations, and other technical inhomogeneities~\cite{bauch_ultralong_2018}. Measurements of the single and double quantum $T_2^*$ in six different locations across the sample yielded average values of $T_2^*\{\text{DQ}\}\!=\!0.70\,(5)\,${\micro}s and $T_2^*\{\text{SQ}\}\!=\!1.12\,(6)\,${\micro}s where the uncertainties indicate one standard deviation. 

Comparison of the single quantum $T_2^*$ and axial-strain-immune double quantum $T_2^*$ provides insight into the dominant dephasing sources across the interrogated ensemble, including the strain inhomogeneity on length scales shorter than the $100\,${\micro}m sample thickness. As expected when limited by magnetic dipolar interactions with the surrounding spin bath, the average $T_2^*\{\text{DQ}\}$ is nearly half the average $T_2^*\{\text{SQ}\}$ due to the effectively doubled gyromagnetic ratio for the double quantum sensing basis~\cite{bauch_ultralong_2018}. These values are consistent with ensemble-NV dephasing dominated by interactions with other NV$^\text{-}$ sensor spins and remaining N$_\text{S}^0$ bath spins, with a residual contribution from strain inhomogeneity across the interrogated volume of approximately 50$\,$kHz. Additionally, these measurements of $T_2^*$ are consistent with batch measurements of the CW-ODMR linewidth, $\gamma$ (see Supplemental Material~\cite{suppl}).

Expanding on the discussion in Sec.~\ref{materialconsiderations}, the product of [NV$^\text{-}$] and $T_2^*$ is used as a material figure of merit to account for the achieved density of [NV$^\text{-}$] sensor spins. In past work~\cite{sturner_compact_2019,grezes_storage_2015}, $^{12}$C-enriched (99.97$\,\%$) HPHT material containing [N$_\text{S}^0$] of $\sim\,$2$\,$ppm as-grown was treated to produce 0.4$\,$ppm [NV$^\text{-}$] and exhibited a $T_2^*$ of $\sim\,$3.2$\,$\micro s. In such samples, the product [NV$^\text{-}$]$\times T^*_2$ is 1.3$\,$\micro s$\cdot$ppm, which compares to 2.7$\,$\micro s$\cdot$ppm for the optimized material characterized in this work. The 2.7$\,$\micro s$\cdot$ppm value also compares favorably to compiled assessments of samples in the literature~\cite{acosta_diamonds_2009,barry_sensitivity_2019,nobauer_creation_2013}.
 
\subsection{NV charge-state and contrast} \label{chargestatecontrast}

Sections \ref{absorptionetc} and \ref{hightanneal} focused on maximizing the value of [N$_\text{S}^0$]/[N$_\text{S}$] ($\chi$), i.e., minimizing charge traps, in as-grown material with the rationale that this would be beneficial to improve the NV charge ratio $\psi$. Hence, it is worthwhile to examine whether the material produced in this study can elucidate the relationship between the concentration of charge-traps in as-grown CVD material and the values of $\psi$ (and ODMR contrast) after irradiation and annealing. 

To demonstrate an understanding and control of charge trap synthesis, a charge-state-detrimental process (based off process P$_1$) was developed that produced [N$_\text{S}$]$\,=\,$17$\,$(1)$\,$ppm with $\chi$ = 0.49$\,$(8) (referred to as process P$_4$), compared to [N$_\text{S}$]$\,=\,$16$\,$(2)$\,$ppm with $\chi$ = 0.81$\,$(2) in process P$_3$. Hence, in this case, the two processes had similar [N$_\text{S}$], but with significantly different levels of acceptors. As expected, the material also had visibly different absorption properties (Sec.~\ref{absorptionetc}) post growth. The degree of variation in N$_\text{S}$ charge fraction $\chi$ was also larger in the case of diamond material with lower $\chi$, implying this process was less controlled.

\begin{table*}[]
\caption{Results obtained from processes P$_3$ and P$_4$, which have similar starting levels of [N$_\text{S}$], but different fractions [N$_\text{S}^0$]/[N$_\text{S}$].  Process P$_3$ was the chosen process for samples reviewed in Sec.~\ref{absorptionetc}-\ref{hightanneal}. Concentrations were determined after exposure to UV.}
\label{tab:p3p4comparison}
\renewcommand{\arraystretch}{1}
\renewcommand{\tabcolsep}{8.5pt}
\begin{tabular}{ccccc}
\hline \hline
\textbf{Process} & \multicolumn{2}{c}{\textbf{As-grown}} & \multicolumn{2}{c}{\textbf{Post irradiation and annealing}} \\ \hline
   & {[}N$_\text{S}${]} (ppm) & {[}N$_\text{S}^0${]}/{[}N$_\text{S}${]} ($\chi$) & NV (ppm)        & {[}NV$^-${]}/{[}NV{]} ($\psi$) \\
P$_3$ & 16$\,$(2) & 0.81$\,$(2)           & 3.8$\,$(2) & 0.62$\,$(5)            \\
P$_4$ & 17$\,$(1) & 0.49$\,$(8)              & 3.6$\,$(1)   & 0.43$\,$(7)              \\ \hline
\end{tabular}
\end{table*}

Three samples grown using process P$_4$ were irradiated to the same dose as that used for process P$_3$ (Sec.~\ref{NVirraddose}) and were annealed utilizing an equivalent profile. The results obtained from these processes are shown in Table~\ref{tab:p3p4comparison}. [NV] and NV charge fraction $\psi$ are reduced in the process P$_4$ sample, demonstrating that grown-in defects in CVD diamond that act as charge acceptors can have a detrimental influence on the properties observed after irradiation and annealing. 

The optical-absorption properties of material grown by these two processes were also investigated. It was found that the P$_4$ material, with a higher starting level of brown coloration (lower $\chi$), still had a higher level of absorption post irradiation and annealing (around 10-15$\,$\% higher in the range 350-550$\,$nm). Increasing [NV], whilst limiting absorption from other defects at wavelengths $<$637$\,$nm, is beneficial from the perspective of reducing absorption from the laser used to excite NV-luminescence; hence, a material with a higher starting $\chi$ is desirable.    

ODMR contrast, which depends upon the NV charge fraction and scales inverse-linearly with magnetic sensitivity, is another critical material-based factor to optimize~\cite{barry_sensitivity_2019,budker_optical_2007,acosta_diamonds_2009}.
The ODMR contrast for material produced using processes P$_4$ and P$_3$ was compared using pulsed-ODMR, as depicted in Fig.~\ref{chargestatecontrast}(a). Measurements were performed as a function of excitation intensity to account for changes in charge state under 532$\,$nm illumination and $T_1$-related effects. A pinhole was introduced to the NV fluorescence collection path of the setup used in Sec.~\ref{strainmitigation} to restrict the collection volume and ensure homogeneous illumination similar to the approach in Ref.~\cite{alsid_photoluminescence_2019}.

\begin{figure}[htbp]
\begin{center}
\begin{overpic}[width=0.95\columnwidth]{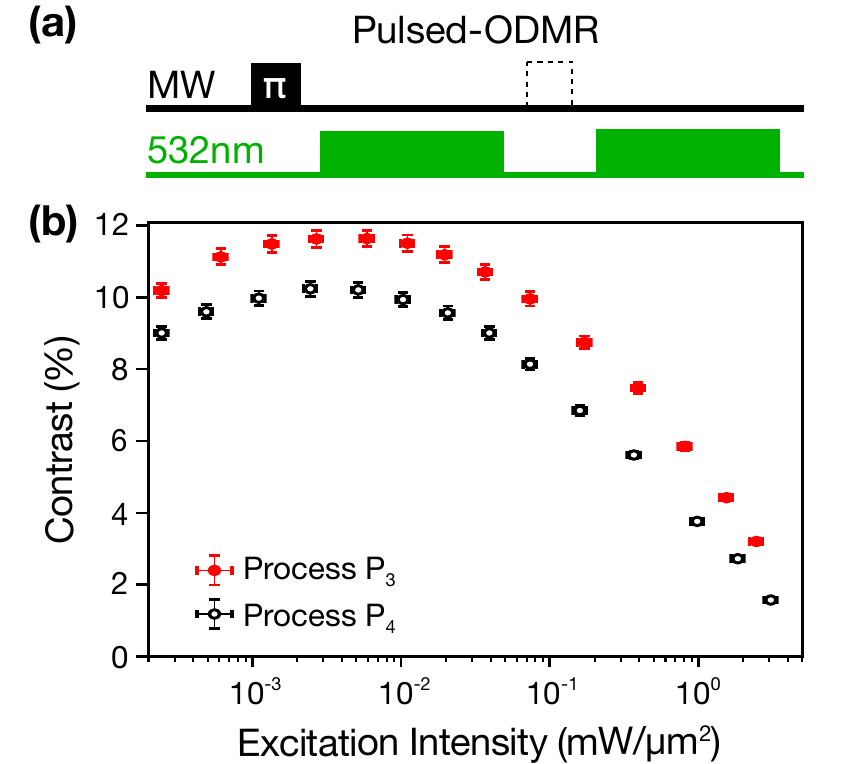}
\end{overpic}
\end{center}
\caption{\label{fig:Contrast} NV pulsed-ODMR contrast measurements of example samples grown with processes P$_3$ and P$_4$. (a) Schematic of the sequence used to measure the NV contrast. Before the first 532$\,$nm optical pulse (green), a resonant microwave (MW) pulse is applied to transfer population from the $m_s\!=\!0$ to the $m_s\!=\!1$ state. The black dashed pulse indicates that no MW pulse was applied before the second optical pulse. Optical pulses are 5$\,$ms in duration and not shown to scale. (b) ODMR contrast as a function of excitation intensity for the two 100$\,$\micro m thick samples produced using process P$_3$ and P$_4$. The reported contrast was determined by the maximum ratio between $n_\text{sig}$/$n_\text{ref}$ during a readout pulse of 532$\,$nm laser light where $n_\text{sig}$ ($n_\text{ref}$) corresponds to the fluorescence measured with(without) an applied MW $\pi$-pulse. Horizontal errors bars indicate an estimated 10$\,\%$ uncertainty in the measured intensity and vertical error bars indicate an estimated 2$\,\%$ uncertainty in measured contrast.}
\end{figure}

Since NV-ensemble devices commonly employ a long-pass filter to partially isolate NV$^\text{-}$ fluorescence from background NV$^0$ fluorescence~\cite{alsid_photoluminescence_2019, aude_craik_microwave-assisted_2018}, a 647$\,$nm long-pass filter was added to replicate realistic experimental conditions. As shown in Fig.~\ref{fig:Contrast}(a), the measured pulsed-ODMR for process P$_3$ exceeds that of process P$_4$ by approximately 20$\%$ across a range of 532$\,$nm excitation intensities spanning from near saturation intensity around 1-3$\,$mW/\micro m$^2$~\cite{barry_sensitivity_2019} (optimal for applications using pulsed measurement protocols) down to 10$^{-4}\,$mW/\micro m$^2$ (similar to the intensities used for CW-ODMR applications~\cite{levine_edlyn_v_principles_2019}). The two samples exhibit maximum contrast for excitation intensities around 5$\times10^{-3}\,$ mW/\micro m$^2$ with values of 12$\,\%$ and 10$\,\%$ for processes P$_3$ and P$_3$, respectively. At higher intensities, the measured contrast decreases for both samples, likely due to reduced NV charge fraction [NV$^\text{-}$]/[NV] with increasing optical intensity~\cite{alsid_photoluminescence_2019,aude_craik_microwave-assisted_2018}. At lower excitation intensities, the measured contrast also decreases because the fraction of NV$^\text{-}$ centers initialized into the $m_s\!=\!0$ state depends upon the ratio of the optical pumping rate to the depolarization rate 1/$T_1$ (see the Appendix of Dréau et al. for further details~\cite{dreau_avoiding_2011}). 

These ODMR contrast measurements on plates produced using processes P$_3$ and P$_4$ further suggest that as-grown defects in CVD diamond that act as acceptors can impact the material properties after irradiation and annealing. Additionally, high initial N$_\text{S}$ charge fraction, $\chi$, appears to be a useful indicator of improved measurement contrast.

\section{Conclusion} \label{conclusion}

This study identifies the important role CVD synthesis parameters can have on the incorporation of unwanted, vacancy-related defects in nitrogen-doped CVD diamond. These defects can act as charge acceptors and likely contribute to the diamond electronic spin bath and are therefore detrimental to sensing applications using ensembles of NV$^\text{-}$ color centers. Crucially, we determine that high [N$_\text{S}^0$]/[N$_\text{S}$] charge fraction $\chi$ can be achieved independent of [N$_\text{S}$] by limiting parasitic defects. Comparison of material grown with the same initial [N$_\text{S}$] before and after irradiation and annealing suggests that improved N$_\text{S}$ charge fraction in as-grown material increases N$_\text{S}\!\rightarrow\,$NV$^\text{-}$ conversion, thereby increasing the density of NV$^\text{-}$ sensor spins and ODMR measurement contrast, both of which benefit sensing applications. In particular, the synthesis processes developed in this work are expected to provide magnetic sensitivity improvements for current NV-ensemble devices without the additional experimental complexity or power consumption associated with advanced spin control or readout techniques~\cite{barry_sensitivity_2019}.

This study also demonstrates the ability to produce ensemble-NV-diamond samples at scale with controlled levels of strain and reproducible [NV$^\text{-}$] and $T_2^*$, shown through characterization of 23 near-identical samples. The NV$^\text{-}$ density was observed to vary by less than 7$\,\%$ with an average of 2.3$\,$(2)$\,$ppm as measured by UV-Vis absorption spectroscopy. Furthermore, birefringence, CW-ODMR stain-imaging, and Ramsey-based $T_2^*$ measurements suggest that careful substrate surface preparation and pre-synthesis etches enable sufficient control over strain inhomogeneity in the material to largely mitigate strain-gradient-induced contributions to ensemble-NV dephasing and limits to application-relevant magnetic field sensitivity.

The correlation of the desired material properties, such as [NV$^\text{-}$] after irradiation and annealing, with simple CVD-growth metrics, such as the color of as-grown samples, enables rapid exploration of large synthesis parameter spaces. This approach provides an efficient framework to develop future diamond material with varying defect densities tailored to specific applications and, beyond exploring growth recipes for other defect densities, future efforts developing micron-scale, NV-rich surface layers based upon the processes demonstrated here will enable advances in NV-ensemble wide-field magnetic imaging applications. For such samples, control of additional qualities such as surface morphology and a well-defined interface between the high-purity diamond substrate and N-doped layer will be critical.

\begin{acknowledgments}
The authors acknowledge Rajesh Patel and Gavin Morley (University of Warwick) for performing the EPR measurements on the as-grown sample examined in Table~\ref{tab:p3characterization}. This material is based upon work supported by, or in part by, the U.S. Army Research Laboratory and the U.S. Army Research Office under Grant No. W911NF-15-1-0548; the National Science Foundation (NSF) Physics of Living Systems (PoLS) program under Grant No. PHY-1504610; the Air Force Office of Scientific Research Award No. FA9550-17-1-0371; the Defense Advanced Research Projects Agency Driven and Nonequilibrium Quantum Systems (DARPA DRINQS) program under Award No. D18AC00033; the Department of Energy (DOE) Quantum Information Science Enabled Discovery (QuantISED) program under Award No. DE‐SC0019396; and Lockheed Martin under Contract No. A32198. Element Six also acknowledges support from the ASTERIQS program, Grant No. 820394, of the European Commission.

\end{acknowledgments}

\appendix

\section{Summary of samples} \label{samplesummary}

Table~\ref{tab:Samples} contains a summary of all samples used in this study including growth process, purpose, and where they are discussed in the text. 

\begin{table*}[h]
\centering
\caption{Summary of diamond samples used in this study}
\renewcommand{\arraystretch}{1.2}
\renewcommand{\tabcolsep}{8pt}
\label{tab:Samples}
\begin{tabular}{c|cccc}
\hline \hline
\textbf{Process} &
  \textbf{\begin{tabular}[c]{@{}c@{}}Purpose of\\  Samples\end{tabular}} &
  \textbf{\begin{tabular}[c]{@{}c@{}}Section of\\  Text\end{tabular}} &
  \textbf{\begin{tabular}[c]{@{}c@{}}Figure/Table \\ Used\end{tabular}} &
  \textbf{Treatment} \\ \hline
\textbf{1} &
  Low $\chi \rightarrow$ Low L$^*$  &
  \ref{absorptionetc} &
  \begin{tabular}[c]{@{}c@{}}Table~\ref{tab:p1p2comparison}, Fig.~\ref{fig:P1P2UVVis}(a), \\ Fig.~\ref{fig:Lightness}, \ref{fig:ChargeNbands}\end{tabular} &
  As-grown \\ \hline
\textbf{2} &
  High $\chi \rightarrow$ High L$^*$ &
  \ref{absorptionetc} &
  \begin{tabular}[c]{@{}c@{}}Table~\ref{tab:p1p2comparison}, Fig.~\ref{fig:P1P2UVVis}(b), \\ Fig.~\ref{fig:P2FTIR}, Fig.~\ref{fig:Lightness}, \ref{fig:ChargeNbands}\end{tabular} &
  As-grown \\ \hline
\multirow{11}{*}{\textbf{3}} &
  \begin{tabular}[c]{@{}c@{}}Evaluate N-related \\ environment \\ (UV-Vis, FTIR, EPR)\end{tabular} &
  \ref{hightanneal} &
  \begin{tabular}[c]{@{}c@{}}Table~\ref{tab:p3characterization}, \\ Table~\ref{tab:1500deganneal}\end{tabular} &
  As-grown \\
 &
  \multicolumn{1}{l}{} &
  \multicolumn{1}{l}{} &
  \multicolumn{1}{l}{} &
  \multicolumn{1}{l}{} \\
 &
  \begin{tabular}[c]{@{}c@{}}High temperature \\ annealing test\end{tabular} &
  \ref{hightanneal} &
  Table~\ref{tab:1500deganneal} &
  1500$^\circ$C anneal \\
 &
  \multicolumn{1}{l}{} &
  \multicolumn{1}{l}{} &
  \multicolumn{1}{l}{} &
  \multicolumn{1}{l}{} \\
 &
  \begin{tabular}[c]{@{}c@{}}Optimize irradiation \\ dose\end{tabular} &
  \ref{NVirraddose} &
  Fig.~\ref{fig:NVirrad}(a,b) &
  \begin{tabular}[c]{@{}c@{}}Irradiated (Variable)\\ and  Annealed\end{tabular} \\
 &
  \multicolumn{1}{l}{} &
  \multicolumn{1}{l}{} &
  \multicolumn{1}{l}{} &
  \multicolumn{1}{l}{} \\
 &
  \begin{tabular}[c]{@{}c@{}}Batch analysis \\ $[$NV$^-]$ and strain \\  \end{tabular} &
  \ref{strainbatch} &
  Fig.~\ref{fig:P3batch} &
  \begin{tabular}[c]{@{}c@{}}Irradiated and \\ Annealed\end{tabular} \\
 &
  \multicolumn{1}{l}{} &
  \multicolumn{1}{l}{} &
  \multicolumn{1}{l}{} &
  \multicolumn{1}{l}{} \\
 &
  Strain shift maps  &
  \ref{strainmitigation} &
  Fig.~\ref{fig:MzMap}(a,b) &
  \begin{tabular}[c]{@{}c@{}}Irradiated and \\ Annealed\end{tabular} \\
 &
  \multicolumn{1}{l}{} &
  \multicolumn{1}{l}{} &
  \multicolumn{1}{l}{} &
  \multicolumn{1}{l}{} \\
 &
  \begin{tabular}[c]{@{}c@{}}High $\chi \rightarrow$ High $\psi$\\ $\rightarrow$ High contrast\end{tabular} &
  \ref{chargestatecontrast} &
  \begin{tabular}[c]{@{}c@{}}Fig.~\ref{fig:Contrast}(b), \\ Table~\ref{tab:p3p4comparison}\end{tabular} &
  \begin{tabular}[c]{@{}c@{}}Irradiated and \\ Annealed\end{tabular} \\ \hline
\textbf{4} &
  \begin{tabular}[c]{@{}c@{}}Low $\chi \rightarrow$ Low $\psi$ \\ $\rightarrow$ Low contrast\end{tabular} &
  \ref{chargestatecontrast} &
  \begin{tabular}[c]{@{}c@{}}Fig.~\ref{fig:Contrast}(b), \\ Table~\ref{tab:p3p4comparison}\end{tabular} &
  \begin{tabular}[c]{@{}c@{}}Irradiated and \\ Annealed\end{tabular} \\ \hline
\end{tabular}
\end{table*}

\clearpage
\bibliography{PurpleDiamondRefs2.bib}
\end{document}